\documentclass{article}

\usepackage{arxiv}

\usepackage[utf8]{inputenc} 
\usepackage[T1]{fontenc}    

\usepackage[british]{babel}
\usepackage{csquotes}

\usepackage{hyperref}       
\hypersetup{bookmarksnumbered}
\hypersetup{breaklinks,colorlinks,linkcolor=blue,citecolor=blue,urlcolor=blue}
\hypersetup{unicode}

\usepackage{booktabs}       
\usepackage{tabularx}       
\usepackage{makecell}	    

\usepackage{siunitx}
\usepackage{microtype}      
\usepackage{symbols}

\usepackage{bibliography}
\usepackage{placeins}
\usepackage{caption}
\usepackage[labelformat=simple]{subcaption}
\usepackage[capitalize]{cleveref}
\crefname{figure}{Figure}{Figures} 

\def\THETITLE{Particle-based plasma simulation using a graph neural network}
\title{\THETITLE}

\date{28th February 2025}

\author{
    \href{https://orcid.org/0000-0003-3587-646X}{\includegraphics[scale=0.07]{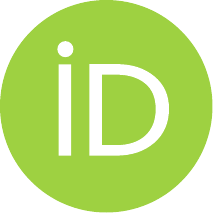}\hspace{1mm}}Marin Mlinarević\textsuperscript{1},
    \href{https://orcid.org/0000-0001-6814-9117}{\includegraphics[scale=0.07]{orcid.pdf}\hspace{1mm}}George K Holt\textsuperscript{2}
    and
    \href{https://orcid.org/0000-0001-9775-0331}{\includegraphics[scale=0.07]{orcid.pdf}\hspace{1mm}}Adriano Agnello\textsuperscript{2} \\
    \\
	{\small \textsuperscript{1}Department of Physics and Astronomy, University College London, Gower Street, London WC1E 6BT, United Kingdom} \\
	{\small \textsuperscript{2}STFC Hartree Centre, Sci-Tech Daresbury, Warrington WA4 4AD, United Kingdom} \\
    \\
    \texttt{marin.mlinarevic.20@ucl.ac.uk, george.holt@stfc.ac.uk, adriano.agnello@stfc.ac.uk}
}

\renewcommand{\shorttitle}{\THETITLE}

\def\THEKEYWORDS{plasma physics, machine learning, graph neural network, particle-in-cell, learned simulator, graph network simulator}
\hypersetup{
pdftitle={\THETITLE},
pdfsubject={physics.plasm-ph, cs.LG, physics.comp-ph},
pdfauthor={Marin Mlinarević},
pdfkeywords={\THEKEYWORDS},
}

\begin{document}
\maketitle

\begin{abstract}
A surrogate model for particle-in-cell plasma simulations based on a graph neural network is presented. The graph is constructed in such a way as to enable the representation of electromagnetic fields on a fixed spatial grid. The model is applied to simulate beams of electrons in one dimension over a wide range of temperatures, drift momenta and densities, and is shown to reproduce two-stream instabilities---a common and fundamental plasma instability. Qualitatively, the characteristic phase-space mixing of counterpropagating electron beams is observed. Quantitatively, the model's performance is evaluated in terms of the accuracy of its predictions of number density distributions, the electric field, and their Fourier decompositions, particularly the growth rate of the fastest-growing unstable mode, as well as particle position, momentum distributions, energy conservation and run time. The model achieves high accuracy with a time step longer than conventional simulation by two orders of magnitude. This work demonstrates that complex plasma dynamics can be learned and shows promise for the development of fast differentiable simulators suitable for solving forward and inverse problems in plasma physics.
\end{abstract}

\section{Introduction}

Plasma simulation is a serious computational challenge with a variety of available methods~\cite{birdsallPlasmaPhysicsComputer2017,dijkPlasmaModellingNumerical2009}.
Particle-in-cell (PIC) is one such popular method that enables the investigation of fundamental plasma processes and applications~\cite{birdsallPlasmaPhysicsComputer2017,epoch,Colonna2017PlasmaMM}. As an established and mature method, modern PIC codes are able to efficiently scale to the largest supercomputers available, which is often required to provide sufficient resolution to simulate real-world phenomena.

This work demonstrates a method for the learned simulation of plasma particle and electromagnetic field dynamics using graph neural networks (GNNs).
We show that constructing the graph representation of the physical domain to include electromagnetic field information is desirable as it enables the study of the effects of fields (both self-consistent and external) on the plasma, which are usually of significant importance for characterizing the complete dynamic behaviour of the system.
In addition to reducing the computation time required for simulations, this kind of learned simulator is differentiable and can therefore also be applied to solve inverse problems by gradient-based optimization, such as automatic design and control~\cite{GNcontrol,inverseDesign,diffusionGenerativeInverseDesign,KumarChoi2023} or physics discovery~\cite{plasma_physics_discovery,discovering_symbolic_models_2020,rediscovering_orbital_mechanics_2023}.

The remainder of this paper is structured as follows. In \cref{sec:related}, we review related works in the field of plasma simulation and machine-learning-based surrogate models for physics simulations which our model is based on. \Cref{sec:simulation} provides an overview of the PIC method for plasma simulation. The dataset used for training and testing our model is described in \cref{sec:data}, and \cref{sec:gns} details the architecture and training procedure. The results of our experiments are presented in \cref{sec:results}. We discuss the performance of the learned simulator in terms of the accuracy of its predictions of particle position and momentum, electric field, the growth rate of two-stream instabilities, energy conservation and run time. Finally, \cref{sec:conclusions} summarizes the conclusions and outlines potential directions for future work towards developing a GNN-based simulator that requires fewer computing resources PIC codes for real-world applications.

\section{Related works}
\label{sec:related}

This work concerns learned dynamic computer simulation of plasma and focuses on the PIC method.
While plasma simulation is a field of research in its own right~\cite{birdsallPlasmaPhysicsComputer2017}, its products are frequently used in numerous fields, including charged particle acceleration~\cite{Tajima2020WakefieldA,Macchi2013IonAB}, plasma photonics~\cite{Lehmann2016TransientPP}, studies of planetary atmospheres~\cite{Birn2012ParticleAI}, nuclear fusion~\cite{Dimits2000ComparisonsAP}, space and astrophysics~\cite{Bchner2003SpacePS}, and others~\cite{Colonna2017PlasmaMM}.

Recent advances in PIC codes have demonstrated high scalability~\cite{DEROUILLAT2018351,10.1145/2503210.2504564,10046112}.
Research has also emerged with a reframing of the PIC algorithm for specific geometric use cases, for example, with azimuthal Fourier mode decomposition~\cite{LIFSCHITZ20091803}, use of spectral methods for the field solver~\cite{GODFREY2014689}, and transforming the simulation space to a Lorentz-boosted frame~\cite{PhysRevLett.98.130405,martins2010exploring}. Monte Carlo methods have been coupled to PIC codes to enable the effects of high-field quantum electrodynamics (QED) on plasma interactions to be studied~\cite{nerush2011laser,ridgers2012dense}.

The ability for neural networks to learn simulation behaviour from training data has been demonstrated in many domains and arises naturally from the universal function approximator property~\cite{HORNIK1989359,bishop1995neural}. An early example applied feedforward neural networks trained with backpropagation~\cite{rumelhart1986learning} to the emulation of various dynamic models~\cite{grzeszczuk1998neuroanimator}. 

Imparting information of the graph structure of the physical space being simulated has been shown to improve learned behaviour by enabling the use of graph neural networks~\cite{1555942,gnn_journal}. For example, Chang et al. demonstrated factorizing a dynamic simulated environment into pairwise interactions between objects enables estimation of a future state of the system by learned composition of the interactions~\cite{Chang2016ACO}. Similarly, Battaglia et al. described interaction networks~\cite{Battaglia2016InteractionNF} and graph networks~\cite{Battaglia2018RelationalIB} for reasoning about objects and their interactions. With their graph network simulator (GNS)~\cite{deepmind}, Sanchez-Gonzalez et al. demonstrated the ability of a graph network in encode-process-decode format to simulate various fluids interacting with rigid bodies, and showed that the learned simulator is able to generalize well beyond the time, space and complexity bounds present in the training data.

A varied body of work exists describing applications of machine learning approaches to PIC methods. Several studies have investigated the replacement of one or several of the field solvers in the PIC loop, including by singular value decomposition for the electric field intensity and magnetic flux density~\cite{Nicolini2019ModelOR}, dynamic mode decomposition for the current density~\cite{Nayak2023AcceleratingPK} and electric potential~\cite{Hesthaven2022AdaptiveSM}, and multilayer perceptron (MLP) and convolutional neural networks for electric field calculation~\cite{Aguilar2021ADL}. Kube et al. trained an MLP to suggest initial solution vectors for a Newton--Krylov solver employed in an implicit PIC algorithm, reducing the required calls to the solver for convergence by 25\%~\cite{Kube2021MachineLA}. MLPs have also been used to replace memory-intensive or expensive-to-evaluate additional modules that are frequently used to extend the PIC loop, such as statistical sampling of Compton scattering~\cite{Badiali2022MachinelearningbasedMI} and lookup tables of pair production cross-sections~\cite{Amaro2024NeuralNS}, both of which have high-field physics applications. In the most application-driven investigations, machine-learning-based surrogate models of end-to-end, laser-plasma-based electron or ion accelerator schemes using MLPs, support vector regressors and Gaussian process regressors have been proposed~\cite{Djordjevi2021ModelingLI,Liu2024DeepLA,Schmitz2023ModelingOA,Desai2023ApplyingMM,Sandberg2024SynthesizingPS,Djordjevi2023TransferLA}. Finally, Carvalho et al. demonstrated the ability of GNS to learn dynamic plasma behaviour from a one-dimensional plasma sheet model, regarded as a precursor to the PIC method~\cite{plasmaGNS}.

\section{Plasma simulation by the particle-in-cell method}
\label{sec:simulation}

The PIC method is one of the most widely used approaches to modelling the dynamic behaviour of plasmas. A thorough overview of the primary algorithm can be found in~\cite{birdsallPlasmaPhysicsComputer2017}. An introductory description is given here for completeness.

Plasma particles are modelled as so-called \textit{superparticles}, which represent a large number of real particles. For example, one superparticle may represent $10^5$ electrons and therefore possess a charge and mass of \qty{1.602e-14}{\coulomb} and \qty{9.109e-26}{\kilo\gram}, respectively. The number of real particles that a superparticle represents is termed its \textit{particle weight}. Superparticles exist in a continuous phase space, and their collective behaviour statistically represents the behaviour of the corresponding real plasma system. A discrete grid is used on which the electric and magnetic fields are defined.

PIC codes contain, at their core, a set of two coupled solvers: the particle pusher and the field solver. In the simplest form of the PIC algorithm, the following steps are repeated to update the superparticles and electromagnetic fields in discrete time:

\begin{itemize}
    \item The electromagnetic field values are interpolated to the positions of each superparticle.
    \item The Lorentz force is solved as the particle equation of motion and used to update the position of each superparticle (the particle push step).
    \item The charge density and current due to particle motion are interpolated onto the field grid.
    \item Maxwell's equations are solved to update the electric and magnetic field values on the grid (the field solve step).
\end{itemize}

The time step must be chosen to satisfy the well-known Courant--Friedrichs--Lewy (CFL) condition~\cite{Courant2015OnTP}: $\Delta t < \Delta x / c$, where $\Delta x$ is the grid resolution and $c$ is the speed of light.

\section{Dataset}
\label{sec:data}

In this section, we describe the specific conditions of the PIC simulations used to generate data for training and testing our learned simulator, and how graphs for input to the GNN were constructed.

\subsection{EPOCH Simulation}
Two collisionless counterpropagating beams of electrons with equal density, temperature, and drift momentum magnitude in a neutralizing background of fixed ions were simulated using the EPOCH PIC code~\cite{epoch} in one spatial dimension. 
Each beam starts with a Gaussian velocity distribution with standard deviation (also known as \textit{thermal velocity}, $\sigma=\vth$) determined by the temperature and mean given by the drift velocity $v_0$.
For sufficiently cold beams, with $\vth/v_0 \lessApprox 0.76$, such a system gives rise to two-stream instability, a well-known phenomenon in plasma physics where the beam particles are bunched periodically in space, with the number density of electrons and therefore the electric field growing exponentially in time until a saturation point. Mathematical derivations and numerical treatments of the two-stream instability under various assumptions are given in \cref{appendix:theory}.

Each beam consists of \num{1600} superparticles, each representing a number of real electrons that depends on the simulated density. The initial distributions of electrons are uniform in space, so that within each example, every superparticle represents the same number of real electrons. 
Periodic boundary conditions were applied with a box length of \qty{500}{km}, meaning particles exiting the box are re-inserted at the opposite boundary with the same velocity.
The electric and magnetic field values were calculated at \num{400} grid points \qty{1250}{m} apart. 
EPOCH uses a time step of $0.95 \Delta x / c$, giving \qty[round-mode=places, round-precision=2]{3.961074008235474}{\micro\second}.

A thousand examples were generated by drawing randomly-sampled temperature values uniformly in the range \qty{0}{K} to \qty{e5}{K}, electron drift momenta from 0 to \qty{5e-24}{kg.m.s^{-1}}, and densities from 5 to 40 electrons per metre per beam. Each simulation was set to evolve for a total time of \qty{0.15}{s}.
For each example, \num{1001} snapshots were recorded approximately \qty{0.15}{ms} apart, corresponding to about \num[round-mode=places, round-precision=1]{37.86851739910307} EPOCH time steps. As the interval of \qty{0.15}{ms} does not match an exact number of internal EPOCH time steps, the time between snapshots turns out to deviate from \qty{0.15}{ms} by up to \qty{3.4}{\micro\second}, which has to be accounted for when calculating input features for and position updates using the GNN simulator.
To train models with a longer time step of \qty{0.60}{ms}, every fourth snapshot from the same sample was used. 940 examples were set aside for training, 30 for validation, and 30 for testing.

\subsection{Graph}
Particle-based simulation naturally lends itself to a graph representation where each simulation particle is represented by a node and edges enable cross-particle interactions.
Nodes can also represent electromagnetic fields at grid points.

Experiments were first conducted to train models using only nodes representing superparticles. Six node features were used: the five most recent velocities (computed from displacements between the latest six snapshots) and particle weights.

Further experiments also trained models with additional nodes representing the electric field at the grid points from the EPOCH simulation, henceforth referred to as \textit{field nodes}. In this case, the length of the input sequence from which velocities are computed was optimized through the procedure described in \cref{appendix:hyperparameter_optimisation}, and additional features were added to all nodes: the component of the electric field in the direction of the motion of the particles in all the snapshots in the input sequence, and node type (0 for grid points and 1 for superparticles). The particle weights were set to 0 for the field nodes, and the electric field was set to 0 for the particle nodes.

Edges were drawn between nodes representing superparticles or grid points within a prescribed connectivity radius.
When there were more than 128 neighbours within this range, only the 128 closest ones were connected. The relative displacement between the superparticles or grid points was encoded into the edge features.

\section{Graph network simulator}
\label{sec:gns}
The models used for this work are based on an implementation of the GNS framework~\cite{deepmind} by Kumar and Vantassel~\cite{pytorchGNS} using PyTorch Geometric~\cite{pytorch,pytorch_geometric}. The principle of the GNS is to predict the acceleration of each particle node, and the field at field nodes, from current and previous node and edge features. Updated positions are then calculated based on the predicted acceleration as described in \cref{position_update}, and a new graph is constructed from these to continue the simulation. We follow the convention described in Ref.~\cite{deepmind} and refer to the result of repeating this procedure as a \textit{rollout}.

\subsection{Graph network architecture}
The model consists of an encoder, processor and decoder. The encoder embeds the node and edge features to latent feature vectors $\vect{x}_i$ and $\vect{e}_{ij}$, respectively, using MLPs.
The underlying architecture of the processor is a message-passing GNN~\cite{message_passing}, where the message passing between nodes on the graph simulates particle interactions. More precisely, in the $m$th message-passing step, the feature vector of the $i$th node, $\vect{x}_{i,m}$, is updated according to the equation:
\begin{equation*}
    \vect{x}_{i,m} = \vect{x}_{i,m-1} + F_m(\vect{x}_{i,m-1}, \sum_j \phi_m(\vect{x}_{i, m-1}, \vect{x}_{j, m-1}, \vect{e}_{ij})),
\end{equation*}
where $F_m$ and $\phi_m$ are MLPs, and $j$ indexes the nodes neighbouring node $i$. The outputs of $\phi_m$ can be thought of as messages, which are aggregated by summation. The $F_m$ network uses these aggregated messages and the current node features to compute updated node features.
Finally, the decoder is an MLP which maps the output latent feature vector into predicted acceleration and electric field values.
The rectified linear unit (ReLU) activation function is used for all hidden layers of the MLPs, and layer normalization~\cite{layernormalization} is used for the outputs of all MLPs except the decoder.

Note the similarity of the message-passing GNN architecture to the PIC loop, particularly when using field nodes, which makes it a good model: 
message passing to particle nodes corresponds to interpolating electromagnetic field values onto the superparticle positions, 
the computation of updated node features using the aggregated messages corresponds to the particle push step, 
message passing from particle to field nodes corresponds to the interpolation of charge and current due to particle motion onto the field grid, 
and message passing between field nodes and the node feature update corresponds to the field solve step.
However, unlike the PIC loop, the GNN can update particle and grid features in parallel and is not subject to the CFL condition, which could allow for faster simulation.

\subsection{Velocity and position update}
\label{position_update}
Given the acceleration $a_n$ at time step $n$, the next position is computed according to the equations:
\begin{equation*}
v_{n+1} = v_n + a_n(t_{n+1}-t_n),
\end{equation*}
and
\begin{equation*}
x_{n+1} = x_n + v_{n+1}(t_{n+1}-t_n).
\end{equation*}
Periodic boundary conditions are implemented by subtracting or adding the box length depending on which side the particle exits.

\subsection{Training and experiments}
For training, the velocities and electric field values used as inputs to the network, as well as target accelerations, are scaled to a mean of 0 and standard deviation of 1, while particle weights are scaled to the range $[0,1]$. The edge feature (the relative displacement of the nodes) is scaled by dividing by the connectivity radius. Random walk noise is added to the input velocities and electric field values by adding a random value drawn from a normal distribution with a mean of zero at each time step in the input sequence and adding it to the value from the previous time step. The standard deviation of the noise is given in \cref{tab:model_parameters}.
The addition of random noise was shown to mitigate the accumulation of error over long rollouts in Ref.~\cite{deepmind}, in which the authors postulate that the network benefits from learning to make predictions from imperfect inputs.

The target acceleration is computed from the positions of the particles in a sequence of three snapshots, according to 
\begin{equation*}
    a = \frac{1}{t_3-t_2}\left(\frac{x_3-x_2}{t_3-t_2}-\frac{x_2-x_1}{t_2-t_1}\right).
\end{equation*}
The network is trained to minimize the mean squared error (MSE) on the predicted acceleration for a single step if field nodes are not used. For models using electric field nodes, the loss function is modified to be the sum of the MSE on the acceleration for particle nodes and the MSE on the electric field for field nodes. The Adam optimizer~\cite{adam} is used with a step-based learning rate decay schedule given by
\begin{equation*}
	\eta_k = \eta_0 d^{\frac{k}{r}},
\end{equation*}
where $k$ is the training step, $\eta_0$ is the initial learning rate, $d$ is the decay factor, and $r$ is the drop rate defining the number of training steps (gradient updates) after which the decay rate should drop to $\eta_0 d$. The values of these parameters are given in \cref{tab:model_parameters}.
To parallelize training across multiple GPUs, different samples are loaded on each GPU, and gradients are computed on each device and averaged before updating the model weights.

Two models were trained without using the electric field feature: model A with a time step of approximately \qty{0.15}{ms} and model B with a time step of approximately \qty{0.60}{ms}, with a batch size of 2, for 4 million training steps on two Nvidia V100 GPUs. Each model took 27 days to train. The model hyperparameters are given in \cref{tab:model_parameters}. The same hyperparameters (number and size of hidden layers) were used for all MLPs in each learned simulator model. The connectivity radius of \qty{7.5}{km} is 15\% of the total box length, the same fraction used by Sanchez-Gonzalez et al. in most physical domains~\cite{deepmind} and corresponds to the length of six grid cells.

\begin{table}[ht]
	\caption{The parameters of and validation loss achieved by three models A, B and C. Models A and B differ only in the time step used in the training data, whereas in Model C the electric field node feature is added and a different set of hyperparameters was found through optimization. The standard deviation of noise given is the value used in the last step of the input sequence, accumulated by adding the variance of the noise at each time step.}
	\centering
	\begin{tabular}{lccc}
		\toprule
		Parameter  & Model A &  Model B & Model C\\
		\midrule
		Time step [\unit{ms}] & 0.15 & 0.60 & 0.60 \\
        Electric field feature & No & No &Yes\\
		Initial learning rate $\eta_0$ & \num{e-4} & \num{e-4} & \num[round-mode=figures, round-precision=3, scientific-notation=true]{0.0009714555159814416}\\
		Learning rate decay factor $d$ & 0.1 & 0.1 & \num[scientific-notation=true, round-mode=figures, round-precision=2]{0.004378300173832424}\\
		Learning rate drop rate $r$ (number of steps) & \num{5e6} & \num{5e6} & \num[scientific-notation=true, round-mode=figures, round-precision=3]{769410.8846882174}\\
		Standard deviation of velocity noise in the last step [\unit{\metre\per\second}] & \num{6.7e-4} & \num{6.7e-4} & \num[scientific-notation=true, round-mode=figures, round-precision=3]{1.1441476875355638e-06}\\
		Standard deviation of electric field noise in the last step [\unit{\newton\per\coulomb}] & - & - & \num[scientific-notation=true, round-mode=figures, round-precision=3]{3.027371262125036e-17}\\
		Number of message-passing steps & 10 & 10 & 11\\
		Number of snapshots in input sequence & 6 & 6 & 8\\
		Connectivity radius [\unit{km}] & 7.5 & 7.5 & \num[round-mode=figures, round-precision=2]{2.532908790395294}\\
		Number of hidden layers in MLPs & 2 & 2 & 1\\
		Number of latent features & 128 & 128 & 209\\
		Number of nodes in each hidden layer of MLPs & 128 & 128 & 185\\
		\midrule
		Minimum validation loss [\unit{km^2}] & \num[scientific-notation=true, round-mode=figures, round-precision=3]{9.184999e3} &
		                                        \num[scientific-notation=true, round-mode=figures, round-precision=3]{7.574420e3} & 
												\num[scientific-notation=true, round-mode=figures, round-precision=3]{8.2651192320e3}\\
		Number of training steps to reach minimum validation loss & \num[scientific-notation=true, round-mode=figures, round-precision=3]{3870000} &
		                                                            \num[scientific-notation=true, round-mode=figures, round-precision=3]{3800000} &
																	\num[scientific-notation=true, round-mode=figures, round-precision=3]{96e4}\\
		\bottomrule
	\end{tabular}
	\label{tab:model_parameters}
\end{table}

\begin{figure}
	\centering
	\includegraphics[width=0.7\linewidth]{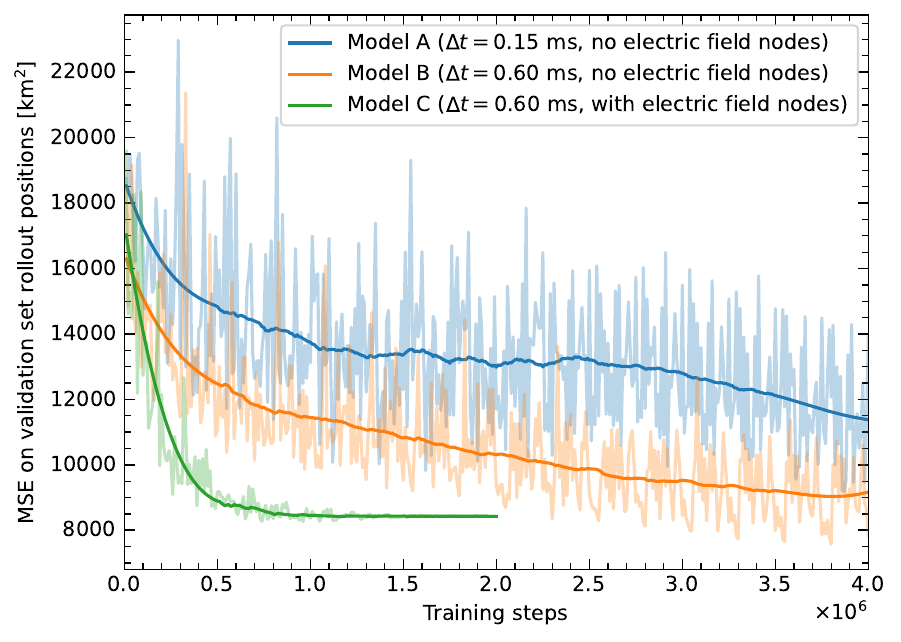}
	\caption{Validation loss curves for training of three models: two without electric field nodes, with time steps of \qty{0.15}{ms} (model A, blue) and \qty{0.60}{ms} (model B, orange), and one with electric field nodes and a time step of \qty{0.60}{ms} (model C, green). The values of the model hyperparameters are given in \cref{tab:model_parameters}. The validation loss is the mean squared error on the predicted position calculated over the full simulation rollout of \qty{0.15}{s}. The faded lines show the loss after every \num{e4} gradient updates, while the solid curves are the result of smoothing by applying a Savitzky--Golay filter~\cite{SavitzkyGolayFilter,SavitzkyGolayFilterErrors} with a cubic polynomial and a window size of 100.}
	\label{fig:validation_loss}
\end{figure}

The MSE on the predicted position calculated over the full simulation rollout of \qty{0.15}{s} was used as the validation loss, and is plotted in \cref{fig:validation_loss}. 
It was computed after every \num{e4} gradient updates.
The validation loss curves still did not plateau after the 4 million gradient updates, which suggests even better performance could be achieved with further training. The model trained with the longer time step is clearly more performant. This is because it takes fewer time steps to compute a rollout simulating the same amount of time, so there is less accumulation of error. This is consistent with the results of Carvalho et al.~\cite{plasmaGNS}, who found that a longer time step resulted in better performance.

Models using field nodes were only trained with the longer time step of \qty{0.60}{ms}. Hyperparameter optimization was required for them to achieve similar performance to model B. Details of the hyperparameter optimization procedure are given in \cref{appendix:hyperparameter_optimisation}. 
The parameters of the model with field nodes achieving the lowest validation loss, model C, are given in \cref{tab:model_parameters}, and its validation loss is plotted in \cref{fig:validation_loss}. 
The training was stopped after 2 million training steps, which was sufficient for the validation loss to plateau in all viable trials with field nodes. 
Training of the best trial lasted 9 days on an Nvidia A100 GPU.
The hyperparameter optimization resulted in a larger learning rate, smaller level of noise added to inputs, greater input sequence length and number of message-passing steps and larger MLP layer sizes, but only one hidden layer instead of two, and the connectivity radius reduced to twice the grid spacing. The added noise for the most performant model found during hyperparameter optimization was set to a negligible level, indicating that adding it did not mitigate error accumulation during rollouts.
For each of the three models A, B and C, the final weights chosen were the ones achieving the lowest validation loss.

\section{Results}
\label{sec:results}
\begin{figure}
	\centering
	\includegraphics[width=\linewidth]{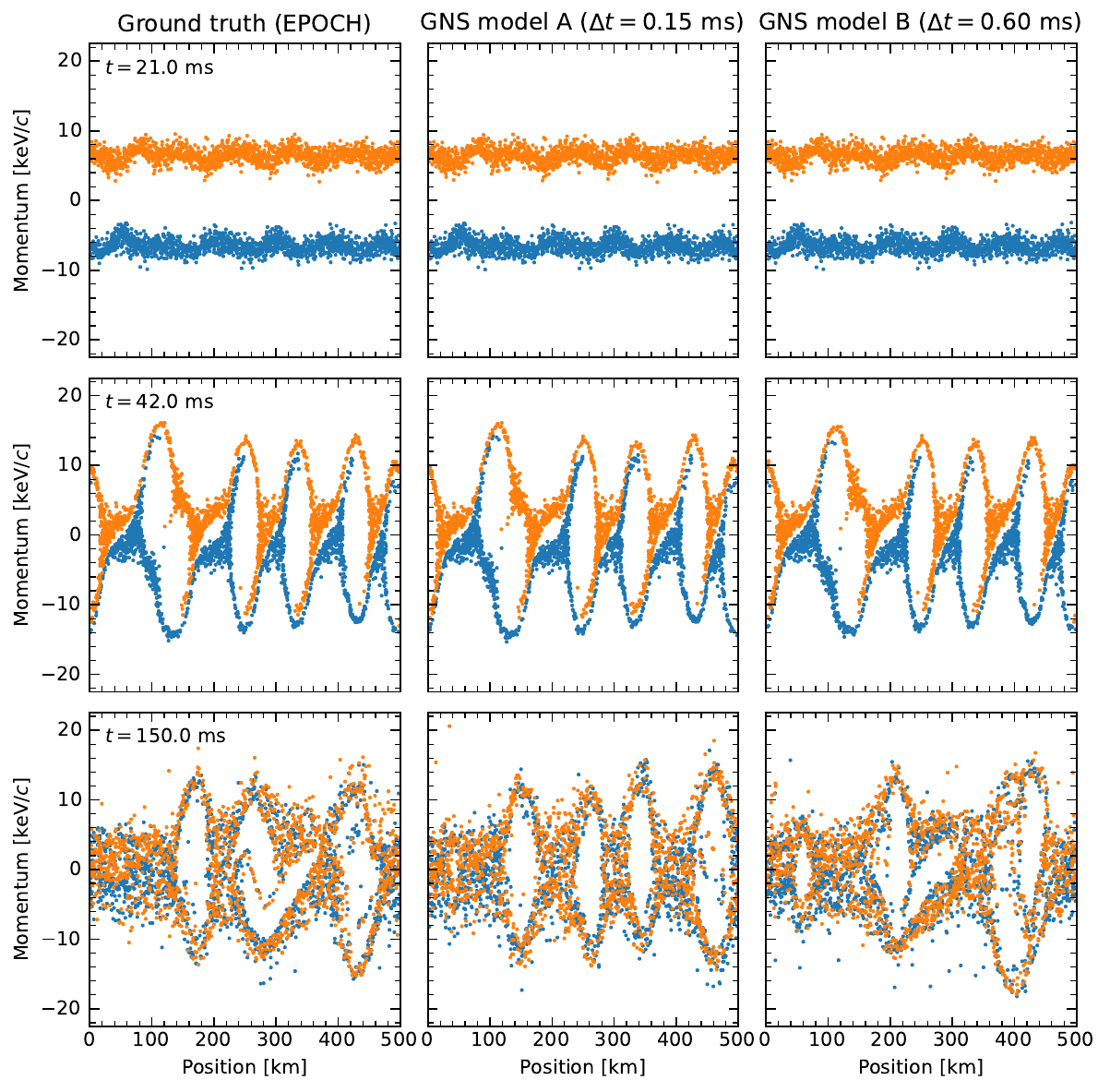}
	\caption{Snapshots from a simulation of two counterpropagating beams of electrons with a density of \num[round-mode=figures, round-precision=2]{24.64989450731967} electrons per metre, initial temperature of \qty[scientific-notation=true, round-mode=figures, round-precision=2]{20363.209227206735}{K}, and drift momentum of \qty[scientific-notation=true, round-mode=figures, round-precision=2]{3.473697853150168e-24}{kg.m.s^{-1}} (giving $\vth/v_0=\num[round-mode=figures, round-precision=2]{0.14030691601102122}$), performed using EPOCH (left column), GNS model A, with a time step of \qty{0.15}{ms} (middle column) and GNS model B, with a time step of \qty{0.60}{ms} (right column). Neither GNS model includes electric field nodes. The values of the model hyperparameters are given in \cref{tab:model_parameters}. The first row shows a plot of electron momentum against position after a simulated time of \qty{21}{ms}, the second row shows the state after \qty{42}{ms}, and the third row after \qty{150}{ms}. Each point represents a superparticle, which represents around \num[round-mode=figures, round-precision=2]{7703.092} electrons. The colour of the points shows which of the two beams the particle comes from. The plots show the development of the two-stream instability.}
	\label{fig:rollout_noField_bestvalMSE_cold_beam_example}
\end{figure}

\begin{figure}
	\centering
	\includegraphics[width=\linewidth]{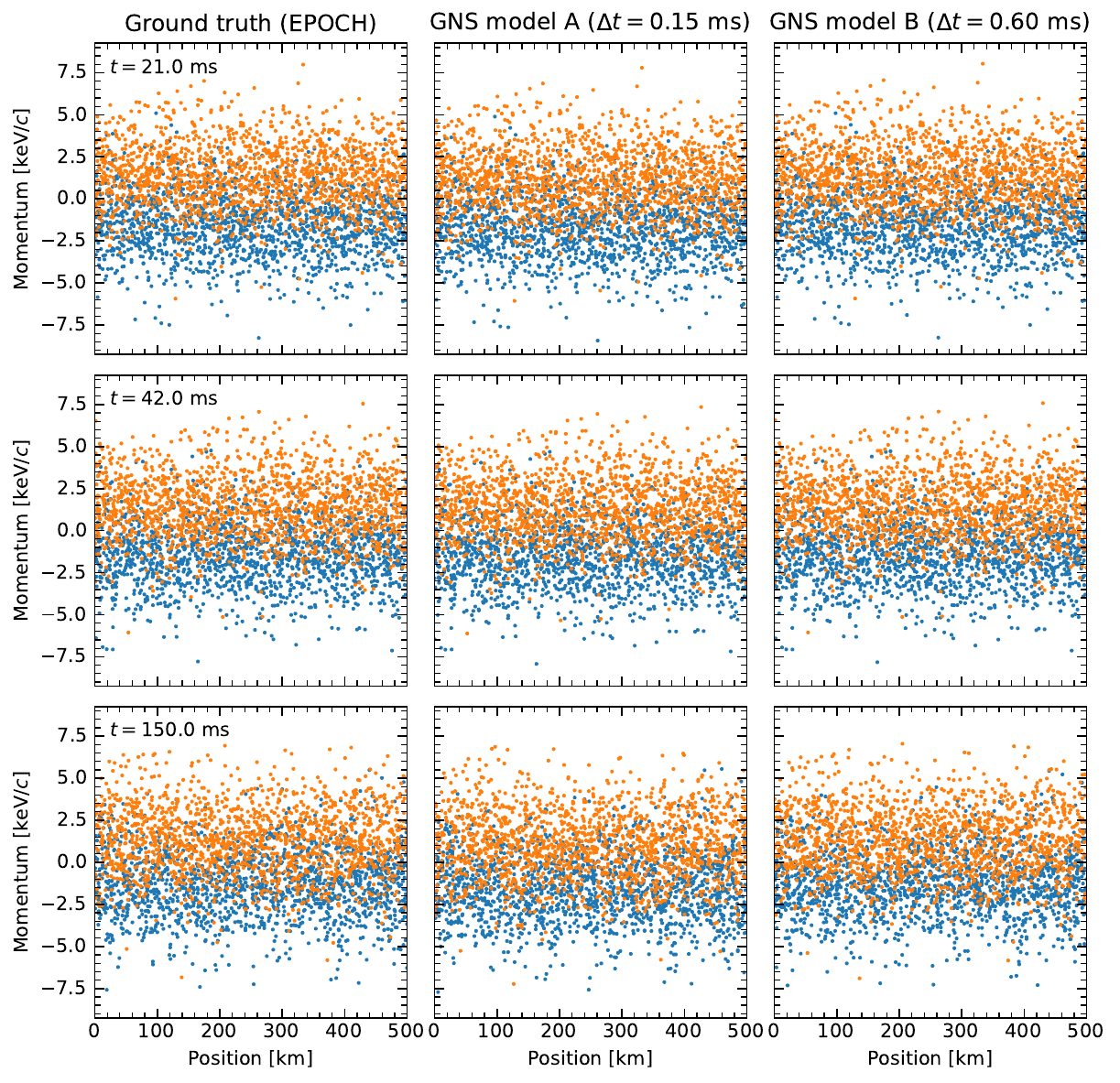}
	\caption{Snapshots from a simulation of two counterpropagating beams of electrons with a density of \num[round-mode=figures, round-precision=1]{5.08246200691531} electrons per metre, initial temperature of \qty[scientific-notation=true, round-mode=figures, round-precision=2]{84259.98604037115}{K}, and drift momentum of \qty[scientific-notation=true, round-mode=figures, round-precision=2]{7.639175348786281e-25}{kg.m.s^{-1}} (giving $\vth/v_0=\num[round-mode=places, round-precision=2]{1.347572224108602}$), performed using EPOCH (left column), GNS model A, with a time step of \qty{0.15}{ms} (middle column) and GNS model B, with a time step of \qty{0.60}{ms} (right column). Neither GNS model includes electric field nodes. The values of the model hyperparameters are given in \cref{tab:model_parameters}. The first row shows a plot of electron momentum against position after a simulated time of \qty{21}{ms}, the second row shows the state after \qty{42}{ms}, and the third row after \qty{150}{ms}. Each point represents a superparticle, which represents around \num[round-mode=figures, round-precision=2]{1588.2694} electrons. The colour of the points shows which of the two beams the particle comes from.}
	\label{fig:rollout_noField_bestvalMSE_low_drift_example}
\end{figure}

\Cref{fig:rollout_noField_bestvalMSE_cold_beam_example} shows a comparison of snapshots from a simulation of two counterpropagating cold beams of electrons (with $\vth/v_0=\num[round-mode=places, round-precision=2]{0.14030691601102122}$) performed using EPOCH and GNS models A and B, which use different time steps, \qty{0.15}{ms} and \qty{0.60}{ms}, respectively, both of which are much greater than the \qty[round-mode=places, round-precision=2]{3.961074008235474}{\micro\second} used internally by EPOCH. This demonstrates that the model can reproduce the two-stream instability, and demonstrates similar fundamental plasma behaviour even with a much lower time resolution. In fact, as expected from the validation loss, model B, with the longer time step, clearly matches the ground truth better than model A at the simulated time of \qty{150.0}{ms}.  \Cref{fig:rollout_noField_bestvalMSE_low_drift_example} shows the GNS also reproduces very similar behaviour to EPOCH in the opposite extreme of warm beams ($\vth/v_0=\num[round-mode=places, round-precision=2]{1.347572224108602}$).

\begin{figure}
	\centering
	\includegraphics[width=\linewidth]{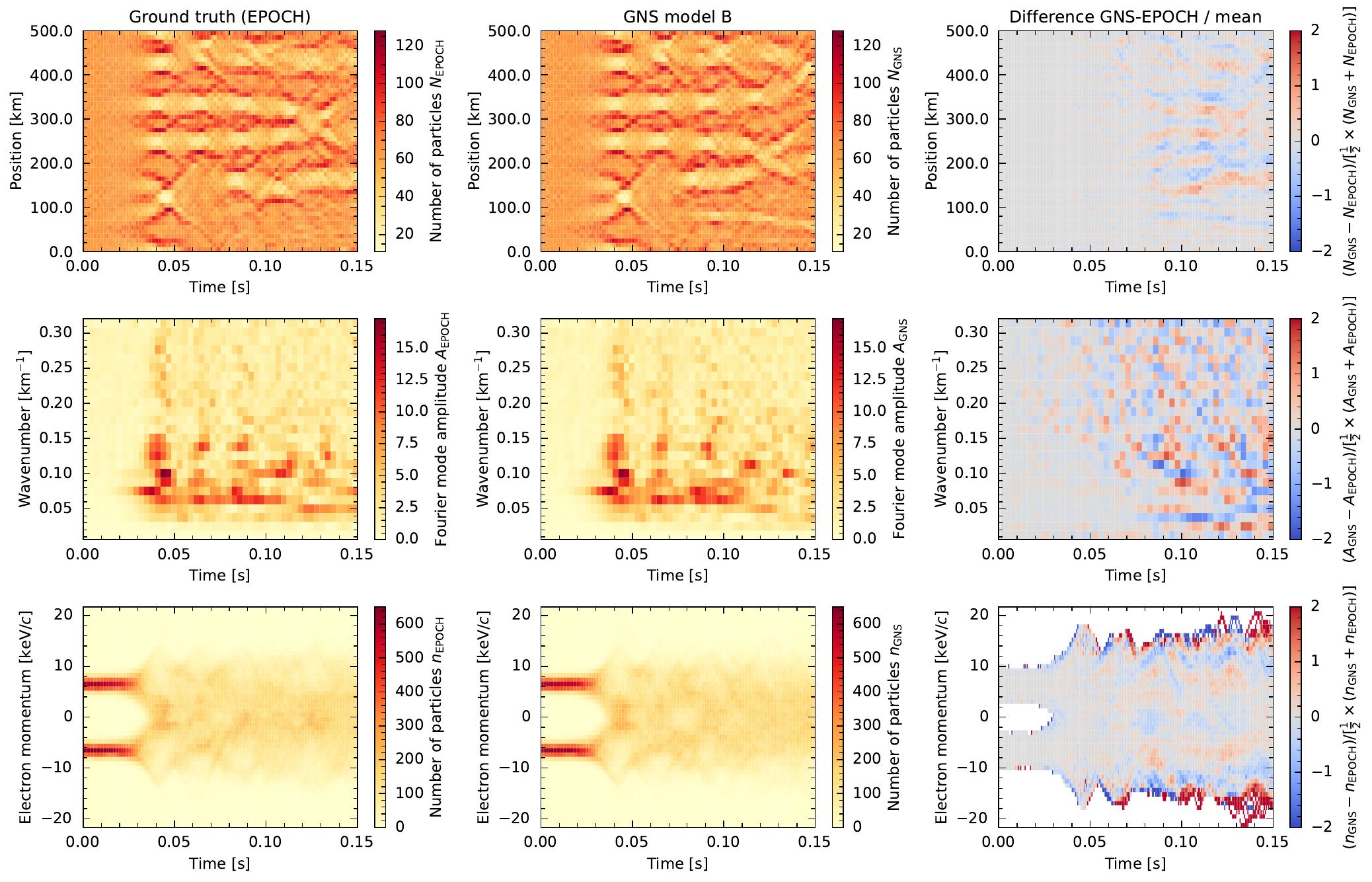}
	\caption{Distribution of the number of superparticles in space over time (top row), its Fourier transform (middle row) and the distribution of electron momenta over time (bottom row) for a simulation of two counterpropagating beams of electrons with a density of 25 electrons per metre, initial temperature of \qty{2.0e4}{K}, and drift momentum of \qty{3.5e-24}{kg.m.s^{-1}} (giving $\vth/v_0=0.14$), performed using EPOCH (leftmost column) and GNS model B (with a time step of \qty{0.60}{ms}, without electric field nodes, in the middle column). The values of the model hyperparameters are given in \cref{tab:model_parameters}. The rightmost column shows the difference between the GNS and EPOCH predictions divided by their mean. The Fourier transform is calculated for each time step and its amplitude averaged over sets of five consecutive time steps. The plots clearly show the development of the two-stream instability.}
	\label{fig:spectrogram_no_field}
\end{figure}

\begin{figure}
	\centering
	\includegraphics[width=\linewidth]{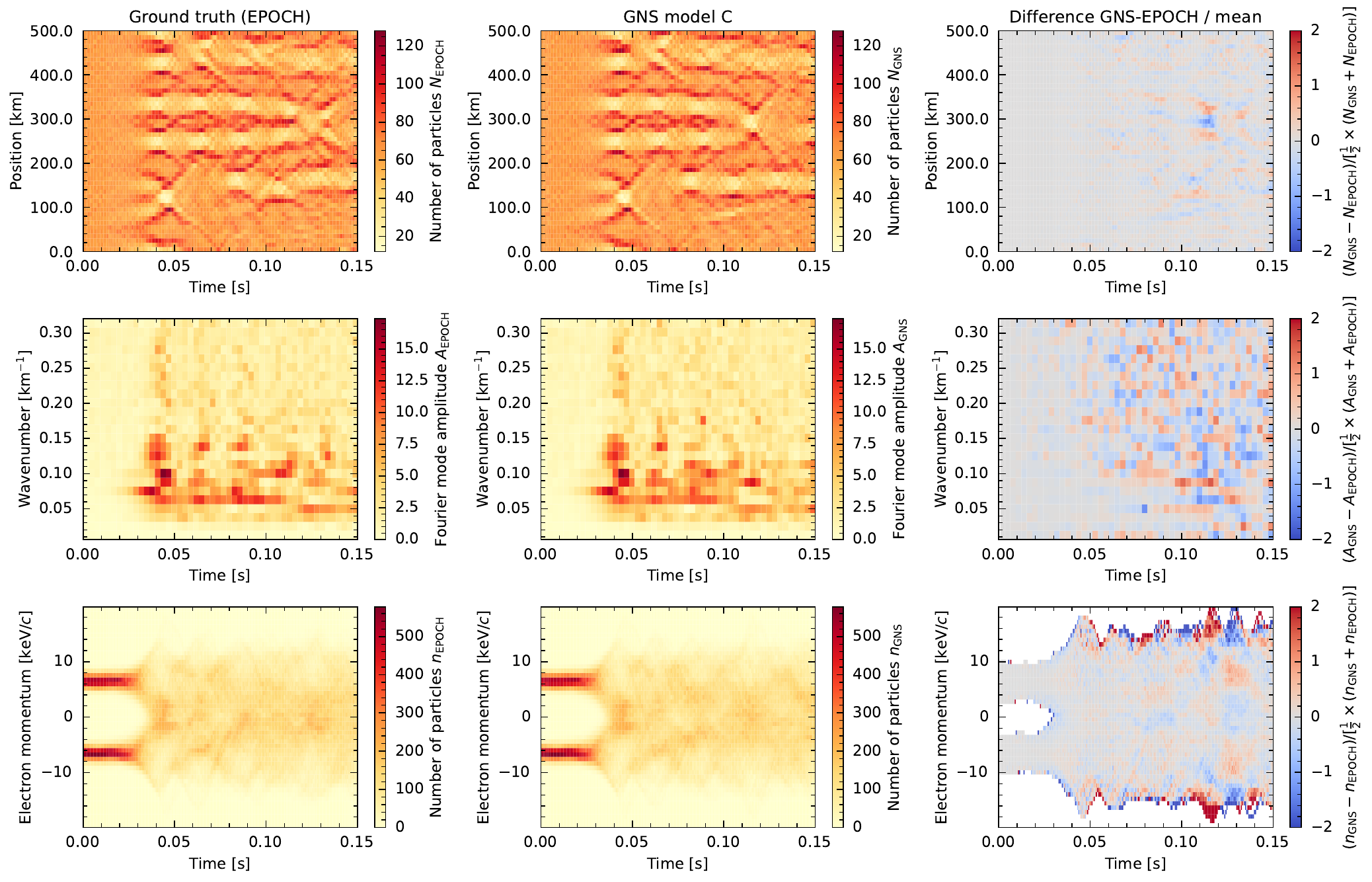}
	\caption{Distribution of the number of superparticles in space over time (top row), its Fourier transform (middle row) and the distribution of electron momenta over time (bottom row), for the same ground truth simulation (leftmost column) as in \cref{fig:spectrogram_no_field}, but performed using GNS model C, which uses field nodes (middle column). The rightmost column shows the difference between the GNS and EPOCH predictions divided by their mean.}
	\label{fig:spectrogram_field}
\end{figure}

\begin{figure}
	\centering
	\includegraphics[width=\linewidth]{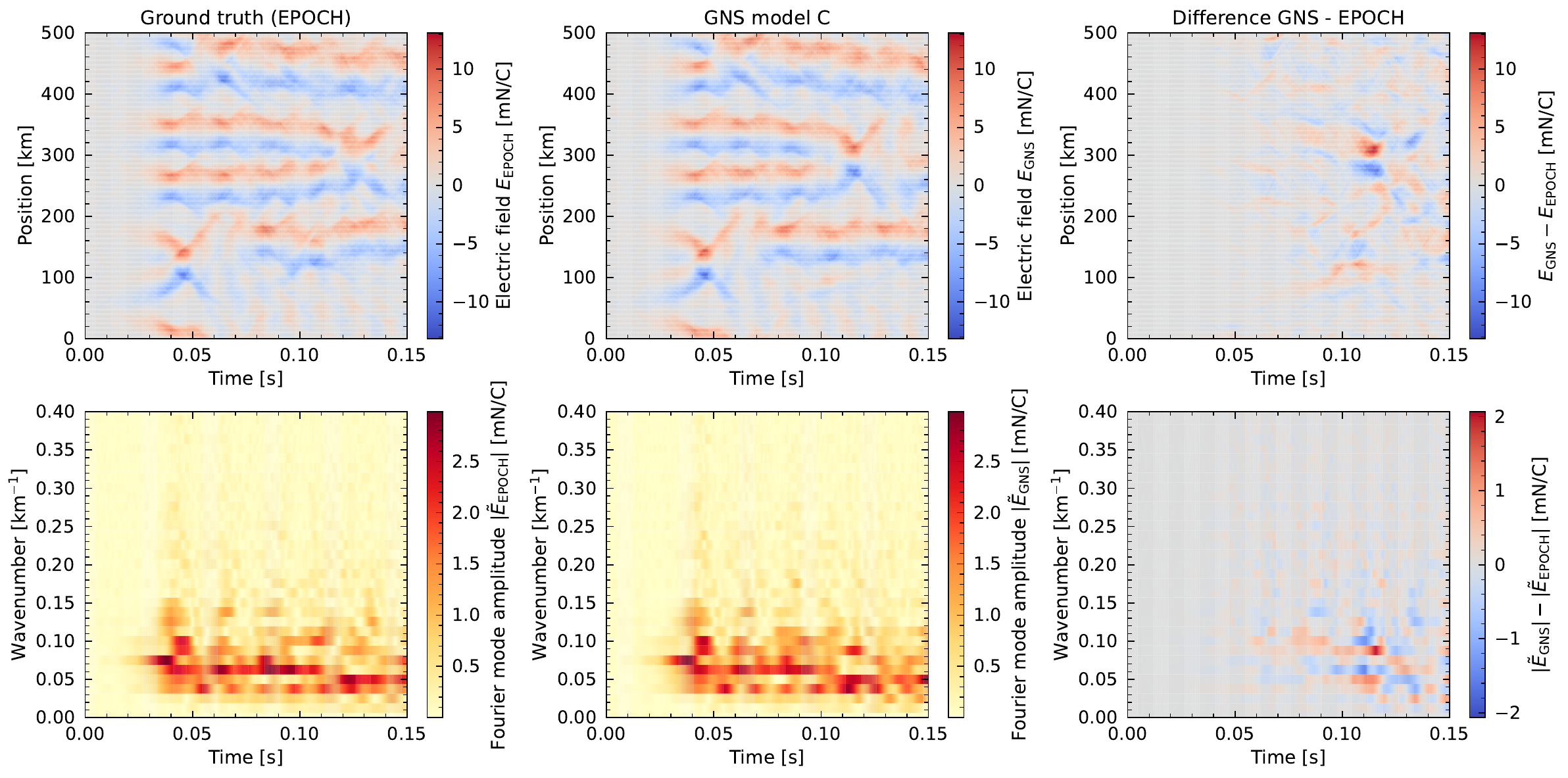}
	\caption{The $x$-component of the electric field over time at 400 grid points in the same simulation presented in \cref{fig:spectrogram_field}, with initial $\vth/v_0=0.14$, performed using EPOCH (left) and GNS model C (middle). The rightmost plots show the difference between the GNS and EPOCH predictions.}
	\label{fig:field_unstable}
\end{figure}

The time evolution of the predicted distributions of particle positions and momenta for the $\vth/v_0=\num[round-mode=places, round-precision=2]{0.14030691601102122}$ case, and the amplitude of the Fourier transform of the position distribution, is shown in \cref{fig:spectrogram_no_field} for model B (without electric field nodes), and in \cref{fig:spectrogram_field} for model C (with electric field nodes). The predicted electric field and the amplitude of its Fourier transform are shown in \cref{fig:field_unstable}, and correspond well to the position distribution. The GNS predictions are generally in good agreement with EPOCH, particularly until \qty{60}{ms}, shortly after the saturation time of the instability.
Notably, model C more accurately predicts the evolution of the number density distribution after about \qty{100}{ms}, in the nonlinear phase, than model B. 
However, since model C is the result of extensive hyperparameter optimization, we make no claim that this is due to the addition of electric field nodes.

\begin{figure}
	\centering
	\includegraphics[width=\linewidth]{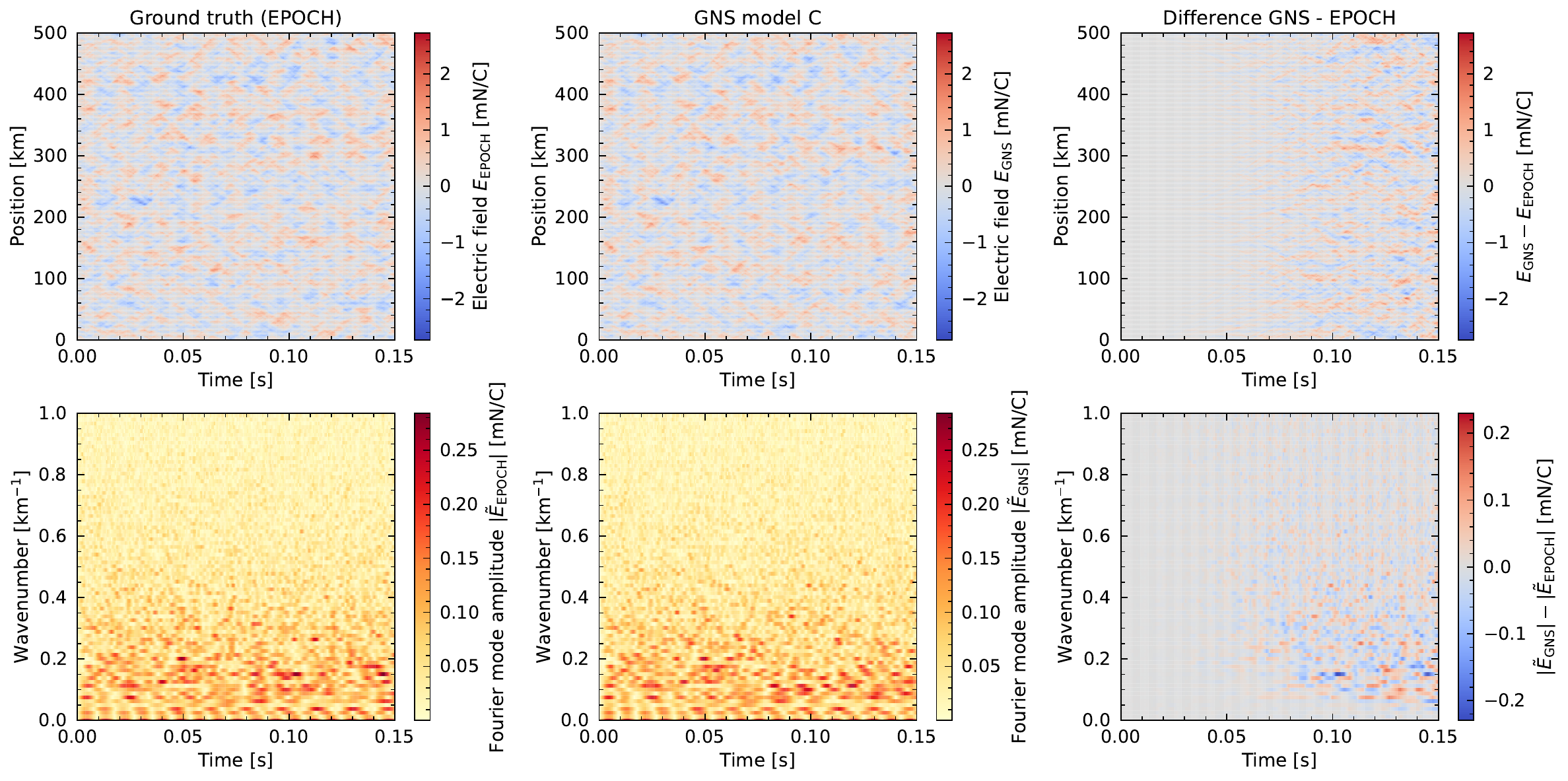}
	\caption{The $x$-component of the electric field over time at 400 grid points in a simulation of two counterpropagating beams of electrons with a density of 19 electrons per metre, initial temperature of \qty[scientific-notation=true, round-mode=figures, round-precision=2]{94518.82186652969}{K}, and drift momentum of \qty[scientific-notation=true, round-mode=figures, round-precision=2]{1.3853422017532118e-24}{kg.m.s^{-1}} (giving $\vth/v_0=0.75$), performed using EPOCH (left) and GNS model C (middle). The rightmost plots show the difference between the GNS and EPOCH predictions. The electric field and its Fourier transform show that there is no instability and no dominant mode, which is in agreement with theoretical predictions.}
	\label{fig:field_stable}
\end{figure}

A quantitative prediction that can be derived analytically in the cold-beam approximation, valid when $\vth \ll v_0$, and by numerical integration in the case of beams with a Gaussian distribution of velocities, is the exponential growth rate of the amplitude of the fastest-growing Fourier mode of the electric field, as explained in \cref{appendix:theory}. This amplitude is plotted in \cref{fig:growth_rates} for EPOCH, GNS and the theoretical prediction, for three examples from the test set with different values of the $\vth/v_0$ ratio from the range where the theory predicts a relatively high instability growth rate. The plots show transient phase, followed by a linear phase where the gradient predicted by the GNS agrees well with EPOCH and the theoretical linear approximation. Beyond this, after the instability saturates, there is chaotic nonlinear phase where exact agreement is not expected because of high sensitivity to noise and initial conditions.
\Cref{fig:field_stable} shows that the GNS also correctly predicts that there is no appreciable growth of instability in an example with $\vth/v_0=0.75$, where theory predicts a very small growth rate.
\begin{figure}
	\centering
	\begin{subfigure}{0.7\textwidth}
		\centering
		\includegraphics[width=\textwidth]{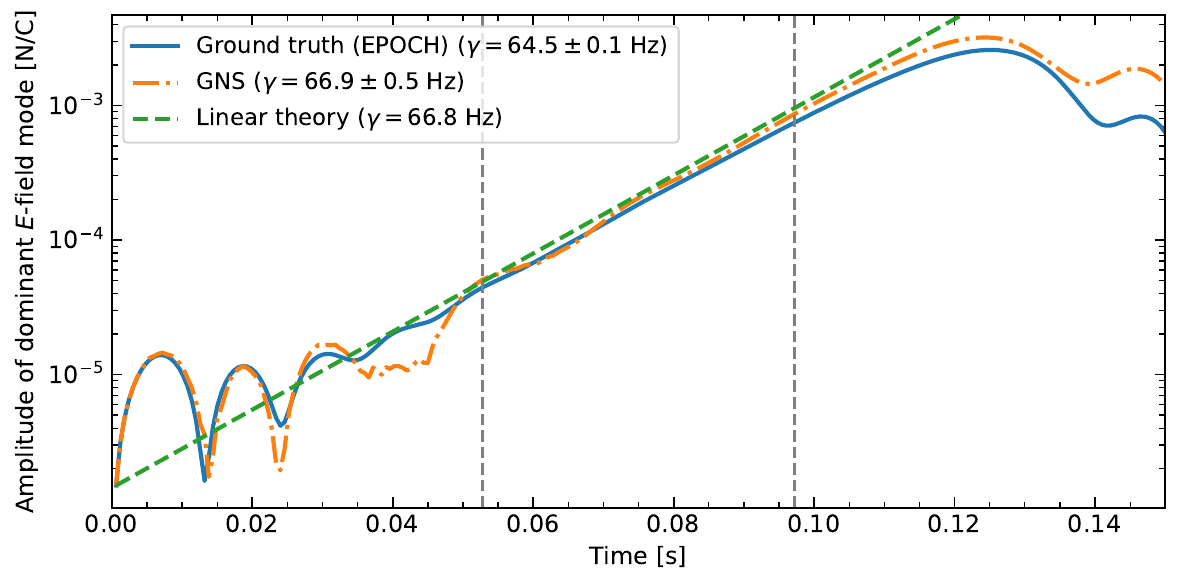}
		\caption{
			$\vth = \qty[round-mode=figures, round-precision=2, scientific-notation=true]{206214.91192064935}{m.s^{-1}}$, 
			$v_0 = \qty[round-mode=figures, round-precision=2, scientific-notation=true]{5418371.830146654}{m.s^{-1}}$,
			$\vth/v_0 = \num[round-mode=places, round-precision=2]{0.03669253607341616}$
		}
		\label{fig:growth_rate_1}
	\end{subfigure}
	\begin{subfigure}{0.7\textwidth}
		\centering
		\includegraphics[width=\textwidth]{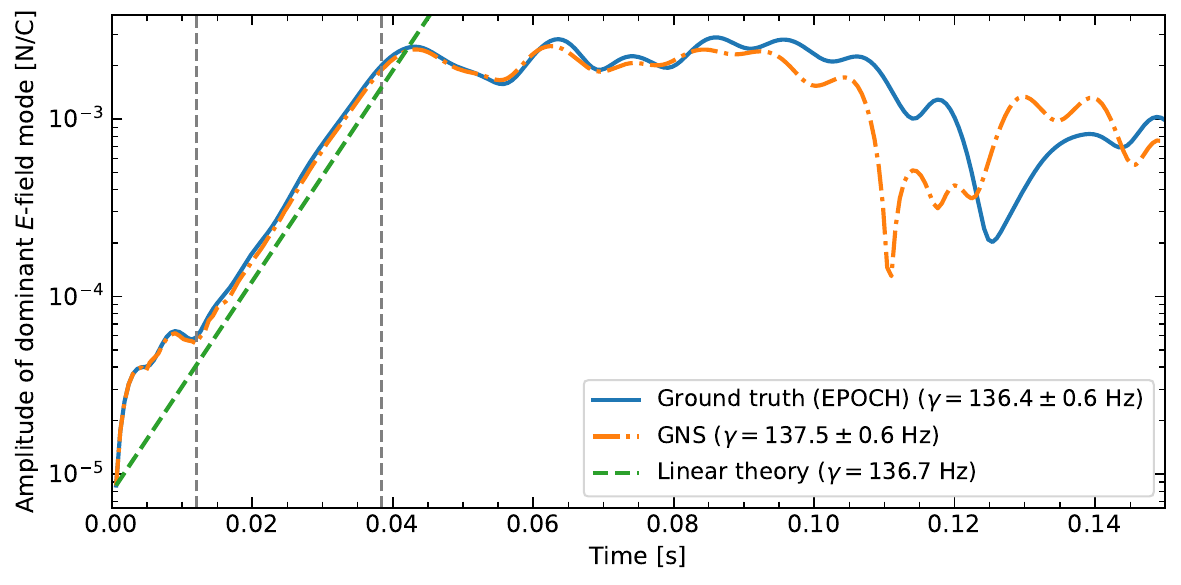}
		\caption{
			$\vth = \qty[round-mode=figures, round-precision=2, scientific-notation=true]{555546.2946479421}{m.s^{-1}}$, 
			$v_0 = \qty[round-mode=figures, round-precision=2, scientific-notation=true]{3813009.7357394476}{m.s^{-1}}$,
			$\vth/v_0 = \num[round-mode=places, round-precision=2]{0.14030691601102122}$
		}
		\label{fig:growth_rate_3}
	\end{subfigure}
    \begin{subfigure}{0.7\textwidth}
		\centering
		\includegraphics[width=\textwidth]{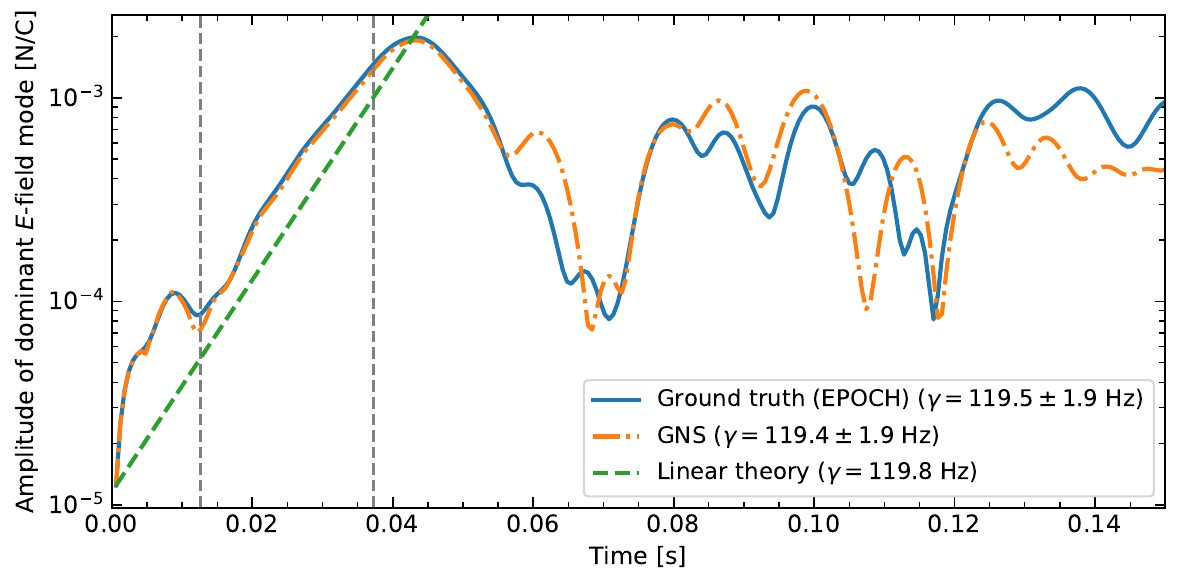}
		\caption{
			$\vth = \qty[round-mode=figures, round-precision=2, scientific-notation=true]{892227.3657809363}{m.s^{-1}}$, 
			$v_0 = \qty[round-mode=figures, round-precision=2, scientific-notation=true]{3697612.568189297}{m.s^{-1}}$,
			$\vth/v_0 = \num[round-mode=places, round-precision=2]{0.2412982294188425}$
		}
		\label{fig:growth_rate_2}
	\end{subfigure}
	\caption{Amplitude of the fastest-growing Fourier mode of the electric field $E$ over time for three simulations with different ratios of initial thermal velocity \vth to drift velocity, $v_0$, as predicted by EPOCH (solid blue line), GNS model C (orange dash-dotted line) and linear perturbation theory (green dashed line), A linear least-squares fit to the data was performed on intervals where the logarithm of the amplitude grows approximately linearly, indicated by grey vertical dashed lines. The best-fit values of the growth rate $\gamma$ are given in the legend. In (a), the analytical cold-beam approximation is valid, whereas in (b) and (c), the theoretical rate computed numerically for Gaussian distributions differs appreciably from the cold-beam approximation.}
	\label{fig:growth_rates}
\end{figure}

\begin{table}[ht]
	\caption{Performance of models A, B and C, defined in \cref{tab:model_parameters}, evaluated on a test dataset, in terms of the mean squared error on the predicted position over all snapshots in a simulation of \qty{0.15}{s} and the execution time of the simulation using an Nvidia V100 GPU. The execution time includes the time taken to produce inputs for the GNS using EPOCH.}
	\centering
	\begin{tabular}{
		l
		S[round-mode=figures, round-precision=3, scientific-notation=true]
        S[round-mode = uncertainty,uncertainty-mode = separate,table-align-uncertainty=true]
	 }
		\toprule
		Model & {MSE on predicted position [\unit{km^2}]} & {Mean execution time [\unit{s}]} \\
		\midrule
		A     & 8276437504e-6 & 131.1(1.6)  \\
		B     & 7377991168e-6 & 35.3(0.2) \\
		C     & 7594137088e-6 & 36.2(1.2) \\
		\bottomrule
	\end{tabular}
	\label{tab:performance_metrics}
\end{table}

\begin{figure}
	\centering
	\begin{subfigure}{0.49\textwidth}
		\centering
		\includegraphics[width=\textwidth]{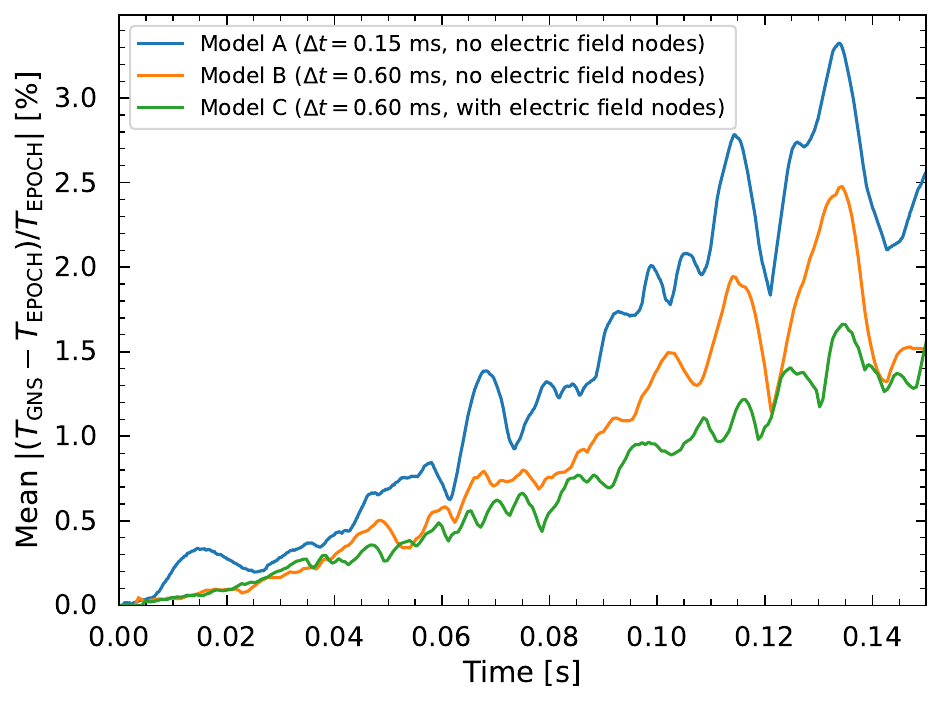}
		\caption{}
		\label{fig:ke_mean_error}
	\end{subfigure}
	\begin{subfigure}{0.49\textwidth}
		\centering
		\includegraphics[width=\textwidth]{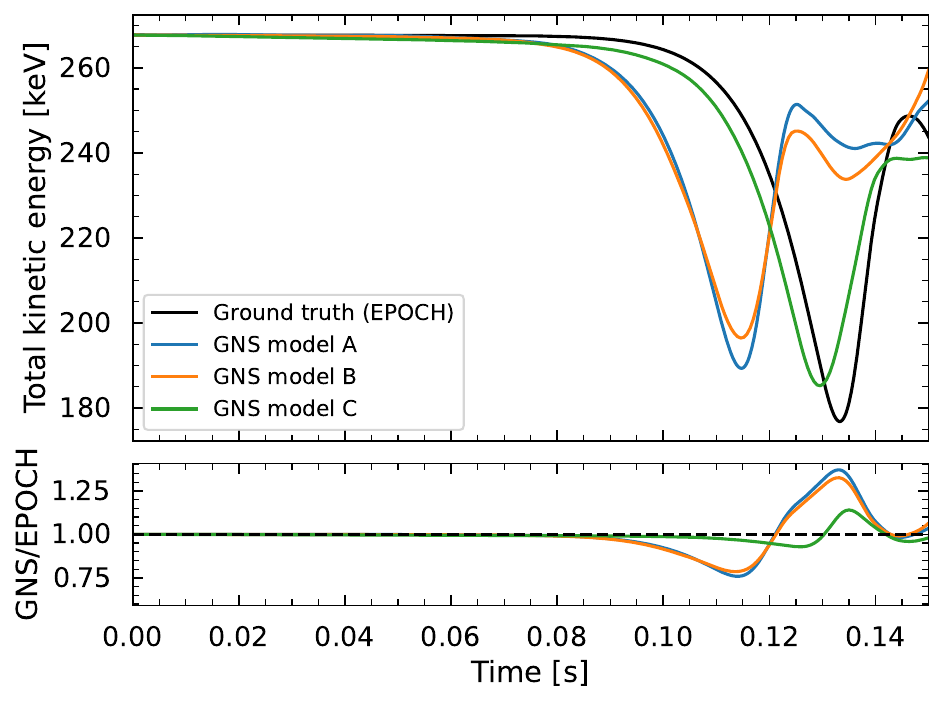}
		\caption{}
		\label{fig:ke_worst_example}
	\end{subfigure}
	\caption{(a) Mean of the absolute values of the relative errors in the total kinetic energy $T$ calculated from predictions of GNS models A, B and C at each time step of a rollout simulating a total time of \qty{0.15}{s}, for simulations in the test sample. 
    (b) The total kinetic energy predicted by EPOCH and GNS models for the example with the largest deviation of the GNS from EPOCH.}
	\label{fig:kinetic_energy}
\end{figure}

\Cref{tab:performance_metrics} summarizes the performance of the GNS models on a test set in terms of the MSE on predicted position over all particles and time steps and the execution time using an Nvidia V100 GPU. Again, models B and C clearly perform better than model A (which uses a shorter time step), achieving an MSE lower by 11\% and 8\%, respectively. Models B and C perform similarly in terms of the MSE on particle position, and on average also in terms of energy conservation, as shown in \cref{fig:ke_mean_error}, which plots the mean error in the total kinetic energy of the particles over time. However, model C (the one with the electric field nodes) produces rollouts with total kinetic energy fluctuations more precisely timed with the ground truth simulations. The temporal departure of these fluctuations for models A and B gives rise to the fluctuations with higher mean error seen in \cref{fig:ke_mean_error}. The largest error in total kinetic energy predicted by model C was 14\%. The total kinetic energy predicted by EPOCH and GNS in the example where this occurs is plotted in \cref{fig:ke_worst_example}, which shows that the discrepancy happens due to the GNS predicting the dip in the total kinetic energy occurring slightly earlier than EPOCH, rather than some wildly different behaviour. 

The mean execution time for the simulations with EPOCH was \qty[round-mode = uncertainty, uncertainty-mode = separate]{39.883708139419554(0.8717084305488936)}{s}. For GNS model A, it was significantly longer at \qty[round-mode = uncertainty, uncertainty-mode = separate]{131.07052267814169(1.6287966711078232)}{s}, while the models using a shorter time step, B and C, were slightly faster at \qty[round-mode = uncertainty, uncertainty-mode = separate]{35.27061470246315(0.17019663729459084)}{s} and \qty[round-mode = uncertainty, uncertainty-mode = separate]{36.2341235365528(1.2025418533825054)}{s}, respectively. The execution time for GNS includes the time taken to produce inputs for the GNS using EPOCH, load the data and write the outputs, but the majority of the time is spent on the running the GNS simulation. However, a proper comparison is difficult because the EPOCH code was written in Fortran, highly optimized, compiled and ran on CPU, whereas the GNS uses GPU acceleration, has received no specific optimization efforts and executes interpreted Python code. Even so, these results demonstrate that the GNS can be sped up significantly by using a longer time step, which sacrifices time resolution but not the accuracy of the prediction of the behaviour of the plasma.

\FloatBarrier

\section{Discussion and conclusions}
\label{sec:conclusions}
The GNS is able to learn simulated dynamic plasma particle and field behaviour to a high level of accuracy.
Over a wide range of plasma conditions, predictions of counterpropagating beam interactions made by the GNS compare very well to ground-truth simulated interactions both quantitatively and qualitatively.
Plasma configurations that give rise to the two-stream instability are well-captured by the GNS, which learns to reproduce the linear and nonlinear phases of the instability.
Growth rates of the GNS-simulated instability match those calculated from linear theory, which shows that the GNS is capable of simulating plasma behaviour to a high level of physical accuracy.

GNS rollouts accumulate errors over time compared to EPOCH simulations.
Adding noise to the inputs during training was not found to reduce this source of error.
By contrast, using larger time steps does reduce error accumulation, as expected. 
There are many potential improvements to the model architecture and training procedure that could be explored to reduce errors, including 
updating edge features with every message-passing step,
optimizing the weighting of the acceleration and field terms in the loss function,
optimizing for whole trajectories instead of single-step predictions,
as well as imposing constraints based on domain knowledge, such as a rotation-equivariant architecture and choosing the connectivity radius based on the Debye length---a distance beyond which electrostatic interactions in a plasma are negligible.

Processing particle-based physics simulation data is memory intensive. As such, available GPU memory is a considerable bottleneck when training GNS models. In our experiments, batch sizes were limited to a maximum of three. It is possible that more performant models could be trained with access to larger GPU memory resources.
Memory requirements could be reduced by overloading the node-level input feature vector components that correspond to velocity for particle nodes and field values for field nodes instead of using zero-padding. However, this could reduce the accuracy of the model as the encoder MLP would have to learn completely different behaviour depending on the node type.

A significant improvement in computation time compared to optimized PIC codes was not achieved, but there is potential for further optimization, both by writing more efficient code, and choosing model parameters, such as the number of message-passing steps, to optimize the trade-off between accuracy and run time, rather than just the accuracy. Furthermore, increasing the time step of the GNS can result in much faster computation at the cost of time resolution; an option that is not available when using conventional simulators, which are bound by more stringent numerical stability criteria. However, the GNS needs to be retrained to use a different time step. It may be possible to avoid this by training a single model using samples with a wide range of time steps and providing this information as a graph-level input feature. The GNS could be used to model PIC simulations with additional modules including, for example, collisions, ionization, or QED effects, which would take longer to run, so there could be a more significant relative reduction in computation time. 

It is straightforward to add additional species by specifying different values for the node-type feature, and to extend the GNS to more dimensions and to include magnetic fields by adding them as additional node features, and adding corresponding terms in the loss function as is done here for the electric field. The GNS could then be used to model interactions with external fields, including lasers, and to tackle inverse design and control problems~\cite{inverseDesign,diffusionGenerativeInverseDesign,KumarChoi2023,laser_plasma_instabilities_in_indirect_drive_inertial_fusion} or physics discovery~\cite{plasma_physics_discovery}.

\section*{Data availability statement}
The trained models described in this paper and the data used to train and evaluate them are openly available at: \url{https://doi.org/10.5281/zenodo.14941475}.

\section*{Acknowledgements}
The authors were supported in part from the STFC International Science Partnerships Fund \textit{ESCAPE} project (ISPF-014).
Marin Mlinarević was supported by the UCL Centre for Doctoral Training in Data Intensive Science (UK STFC Training Grant No. ST/P006736/1).
Computing resources were provided by the STFC Scientific Computing Department's SCARF cluster.
The EPOCH code used in this work was in part funded by the UK EPSRC grants EP/G054950/1, EP/G056803/1, EP/G055165/1, EP/ M022463/1 and EP/P02212X/1.

\appendix
\numberwithin{figure}{section}
\numberwithin{equation}{section}
\numberwithin{table}{section}

\section{Appendix: Dispersion relations and growth rates}
\label{appendix:theory}

In a plasma described by a sum of distribution functions of velocities, $n_{j}f_{j}(v)$, where $n_j$ is the number density and $\int_{-\infty}^\infty f_{j}(v)\mathrm{d}v=1$, linear perturbation theory can be used to show that electrostatic waves of the form
\begin{equation*}
    E = E_0 \mathrm{e}^{\mathrm{i}(kx-\omega t)}
\end{equation*}
develop, where $k$ is the wavenumber, $\omega$ is the angular frequency of the wave, and $x$ and $t$ are position and time coordinates, respectively. These waves satisfy the dispersion relation
\begin{equation}
1\ =\ \sum_{j}\frac{\omegapj^{2}}{k^{2}}\int_{L} \frac{\partial_{v}f_{j}}{v-\omega/k}\ \mathrm{d}v\ \equiv \sum_{j} \epsilon_{j}(\omega,k),
\label{eq:dispersion}
\end{equation}
where $\omegapj$ is the plasma frequency of the species and $L$ denotes the Landau contour passing below the pole at $v=\vph\equiv\omega/k$, the phase velocity of the wave. \Cref{eq:dispersion} is in terms of the one-dimensional velocity distribution (relevant to this paper), and is valid for 3D electrostatic waves by positing $f(v)=\int f_\mathrm{3D}(\vec{v})\delta(\vec{v}\cdot\hat{\mathbf{n}}-v) \mathrm{d}^{3}v,$ where $\hat{\mathbf{n}}$ is the direction of wave propagation~\cite{Krall}. The plasma frequency is given by
\begin{equation*}
    \omegapj = \sqrt{\frac{n_je_j^2}{\epsilon_0m_j}},
\end{equation*}
where $e_j$ is the particle charge, $m_j$ is the particle mass and $\epsilon_0$ is the permittivity of free space.
In general, solutions for $\omega$ can have a real part, which is the angular frequency of a wave oscillating in time, and an imaginary part, which corresponds to Landau damping if negative, or instability (exponential growth) if positive. 
From the linear kinetic theory of perturbations, the dielectric integrals in \cref{eq:dispersion} should be computed assuming $\mathrm{Im}(\omega)>0$ and continued analytically on the complex plane~\cite{Krall}.

In this Appendix, we provide analytical expressions for two-stream systems with different distribution functions, and provide a useful formula to derive growth rates. A special class of distribution functions, described below, can give dispersion relations with purely real solutions in some regimes. However, they all also have unstable regimes with purely imaginary solutions, as they can have a `hole' between two beams in velocity space (which is always unstable as per the Penrose criterion~\cite{Penrose1960}). 

\subsection{Cold two-stream limit}

The cold-stream limit, where $f_{j}=\delta(v-v_{0,j}),$ has a simple analytical form:
\begin{equation*}
1\ =\ \sum_{j} \frac{\omegapj^{2}}{(\omega-v_{j}k)^{2}}\ .
\end{equation*}
For two cold electron streams with initial velocities $\pm v_{0}$ and each with plasma frequency $\omegapeb,$ and a uniform non-moving ion background, the dispersion relation can be derived as
\begin{equation*}
1\ =\ \frac{\omegapeb^{2}}{(\omega-v_{0}k)^{2}}+\frac{\omegapeb^{2}}{(\omega+v_{0}k)^{2}}\ = 2\omegapeb^{2}\frac{\omega^{2}+v_{0}^{2}k^{2}}{(\omega^{2}-v_{0}^{2}k^{2})^{2}}
\end{equation*}
and can be solved explicitly in $\omega^{2}.$ Posing 
\begin{equation*}
\hat{\omega}_{\pm}^{2}\ =\ \left| \hat{k}^{2}+1 \pm \sqrt{1+4 \hat{k}^{2}} \right|
\end{equation*}
with $\hat{k}=k v_{0}/\omegapeb,$ there is always a real solution corresponding to an oscillation in time with $\omega/\omegapeb=\pm\hat{\omega}_{+},$ and one with $\omega/\omegapeb=\pm\hat{\omega}_{-}$ for $k v_{0}/\omegapeb>\sqrt{2}$. For $k v_{0}/\omegapeb<\sqrt{2}$, there are imaginary solutions $\omega/\omegapeb=\pm\mathrm{i}\hat{\omega}_{-}$. The solution with a positive imaginary part corresponds to a field that is growing exponentially in time with no oscillations:
\begin{equation*}
    E = E_0 \text{e}^{\gamma t} \text{e}^{\text{i}kx},
\end{equation*}
with growth rate $\gamma=\hat{\omega}_{-}\omegapeb$. In particular, the fastest-growing mode has growth rate $\gamma=\omegapeb/2$ at $k v_{0}/\omegapeb=\sqrt{3}/2.$ We note that these numbers are the same as in Ref.~\cite{plasmaGNS}, once we account for the fact that the authors express the individual-beam plasma frequencies as $\omegapeb=\omegape/\sqrt{2}$ -- i.e. \omegape as the one from the total density of electrons.

\subsection{Top-hat two-stream behaviour}

The cold-stream case reminds us that the growth rates are set by the only dimensional frequency of the system (\omegape), and that unstable perturbations arise at wavelength larger than $v_{0}/\omegape.$ As the dielectric function in \cref{eq:dispersion} diverges at the resonances $\vph=\pm v_{0},$ four real roots (i.e. stable oscillations) are available whenever $\sum_j\epsilon_j(\omega=0,k)<1.$

A simple example of dispersion relation can be obtained when each beam has a `top-hat' compact support distribution function, constant over ${v_{j}-\sigma <v<v_{j}+\sigma}$ and zero otherwise, yielding
\begin{equation*}
\epsilon_{j}(\omega,k)=\frac{\omegapj^{2}}{(\omega-k v_{j})^{2}-k^{2}\sigma^{2}}\,.
\end{equation*}
For two beams at $\pm v_{0},$ each with plasma frequency $\omegapeb,$ the dispersion relation has the same behaviour as for the cold case with the only change being that
\begin{equation*}
\hat{\omega}_{\pm}^{2} = \left| \hat{k}^{2}(1+\alpha^{2})+1 \pm \sqrt{ 1+ 4\hat{k}^{2}+4\hat{k}^{4}\alpha^{2} }\right|\,,
\end{equation*}
where $\alpha = \sigma/v_{0},$ and the two-stream instability arises for $k^{2}<2\omegapeb^{2}/(v_{0}^{2}-\sigma^2)$ whenever $\alpha<1.$ When $\sigma\geq v_{0},$ the two beams merge into one centred at zero velocity, and two oscillatory solutions arise. When $\alpha^{2}<1,$ the maximum growth rate of instabilities occurs at
\begin{equation*}
\hat{k}_{\star}^{2}=\frac{(1+\alpha^{2})/\sqrt{1-\alpha^{2}}\ -1}{2\alpha^{2}}\,.
\end{equation*}

\subsection{Warm beams with compact support}
The case above can be generalized to other distribution functions with compact support, by formally expanding the denominator in the dielectric integral in \cref{eq:dispersion}. In particular, supposing that $f_{j}(v)=n_{j} \hat{f}(v-v_{0,j})$, then
\begin{eqnarray*}
\nonumber    \epsilon_{j}(\omega,k) & = &
    \frac{\omegapj^{2}}{k^{2}}\int_{L}\hat{f}(u)\frac{\mathrm{d}u}{(v_{0,j}-\vph+u)^{2}} \ =\  \frac{\omegapj^{2}}{k^{2}}\int_{L}\frac{\hat{f}(u)}{(v_{0,j}-\vph)^{2}}\sum_{m\geq0}(-1)^{m}(m+1)\frac{u^{m}\mathrm{d}u}{(v_{0,j}-\vph)^{2}}\\
    & = &  \frac{\omegapj^{2}}{k^{2}}\sum_{m\geq0}(-1)^{m}(m+1)\frac{\mu_{m}}{(v_{0,j}-\vph)^{m+2}}
\end{eqnarray*}
whenever $\vph$ is outside the support of $f,$ and where $\mu_{m}=\int \hat{f}(u)u^{m}\mathrm{d}u/\int \hat{f}(u)\mathrm{d}u.$ If the series expansion converges, then it can also be analytically continued for $\vph=\omega/k$ within the support of $f.$
As an example, for a beam with distribution $\hat{f}\propto (1-|u/\sigma_{j}|)\mathbf{1}_{(-\sigma_{j}<u<\sigma_{j})},$ the dielectric integral series can be resummed as
\begin{equation*}
\epsilon_{j}(\omega,k)=\frac{\omegapj^{2}}{\sigma_{j}^{2}k^{2}}\ln\left( 1+\frac{\sigma_{j}^{2}}{(v_{0,j}-\vph)^{2}} \right)\ .
\end{equation*}
For two beams of this kind, the dispersion relation then becomes:
\begin{equation}
    \frac{2(\hat{\omega}^{2}+\hat{k}^{2})+\hat{k}^{2}\alpha^{2}}{(\hat{\omega}^{2}-\hat{k}^{2})^{2}}\ =\ \frac{\mathrm{e}^{\sigma_{0}^{2}k^{2}/\omegapeb^{2}}-1}{\sigma_{0}^{2}k^{2}/\omegapeb^{2} }\ \equiv\  1+\frac{1}{2}\alpha^{2}\hat{k}^{2}b_{0}^{2}\,,
    \label{eq:spiky}
\end{equation}
with similar properties as the ones seen for the case of top-hat beams. Expanding around $\alpha=0,$ unstable modes occur at $\hat{k}^{2}\leq 2-\alpha^{2}/2,$ and the instability disappears once $|\alpha|\geq 2.$ This coincides with a Newcomb-stable distribution function ($v\partial_{v}f\leq0$ everywhere).

\subsection{Beams with non-compact support}

In the case of two streams with Gaussian distributions $f_{\pm}(v)\propto \mathrm{e}^{-(v\mp v_{0})^{2}/2\sigma^{2}},$ one may formally expand the principal-value integral to obtain
\begin{equation*}
\epsilon_{j}(\omega,k)=\sum_{m}\frac{\omegapj^{2}k^{2m}\sigma_{j}^{2m}(2m+1)(2m-1)!!}{(kv_{0,j}-\omega)^{2m+2}}\,,
\end{equation*}
which however is a divergent series. In the previous examples, the series expansion converges for beam distribution functions with compact support at all values of $\vph=\omega/k$ where $f(\vph)=\partial_{v}f(\vph)=0,$ and has an analytical continuation inside the beam support. For a Gaussian distribution, the expansion is typically truncated to the first two orders, to obtain the Bohm--Gross dispersion relation of Langmuir waves ($\omega^{2}=\omega_\text{p}^{2}+3\vth^{2} k^{2}$) and of the gentle-bump instability~\cite{Krall}, under the assumption $\sigma\ll |\vph-v_{0}|.$ In our case:
\begin{align}
\nonumber 1 & = \sum_{b}\frac{\omegapeb^{2}}{k^{2}} \int_{L}\frac{\partial_{v}f}{v-\omega/k}\mathrm{d}v\approx
\sum_{b}\frac{\omegapeb^{2}}{k^{2}} \int_{L}\left(\frac{u^{2}}{(v_{b}-\omega/k)^{2}} +\frac{u^{4}}{(v_{b}-\omega/k)^{4}}+...\right)\frac{\mathrm{e}^{-u^{2}/2}}{\sqrt{2\pi}}\mathrm{d}u\\
  & \approx  \frac{2\omegapeb^{2}}{k^{2}} \left( \frac{v_{0}^{2}+(\omega/k)^{2}}{(v_{0}^{2}-(\omega/k)^{2})^{2}}+3\sigma^{2}\frac{(1-6(\omega/k v_{0})^{2}+(\omega/k v_{0})^{4})}{(1-(\omega/k v_{0})^{2})^{4}} \right)\ +\ \mathcal{O}(\sigma^{4}/v_{0}^{4})\,. \label{eq:asympgauss}
\end{align}

\begin{figure}
	\centering
	\begin{subfigure}{0.49\textwidth}
		\includegraphics[width=\textwidth]{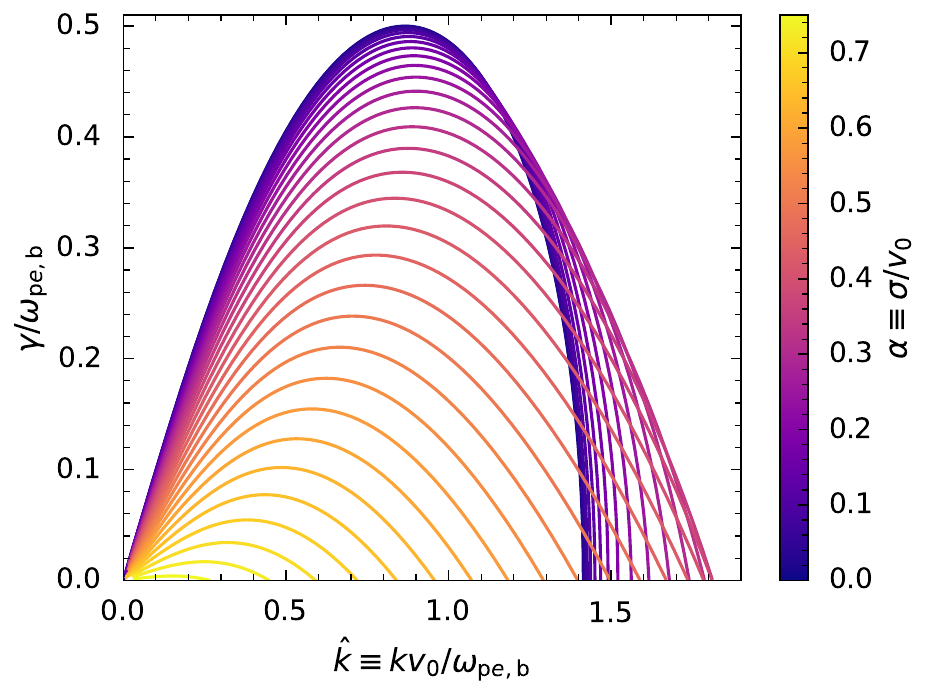}
		\caption{}
		\label{fig:gamma_vs_k}
	\end{subfigure}
	\begin{subfigure}{0.49\textwidth}
		\includegraphics[width=\textwidth]{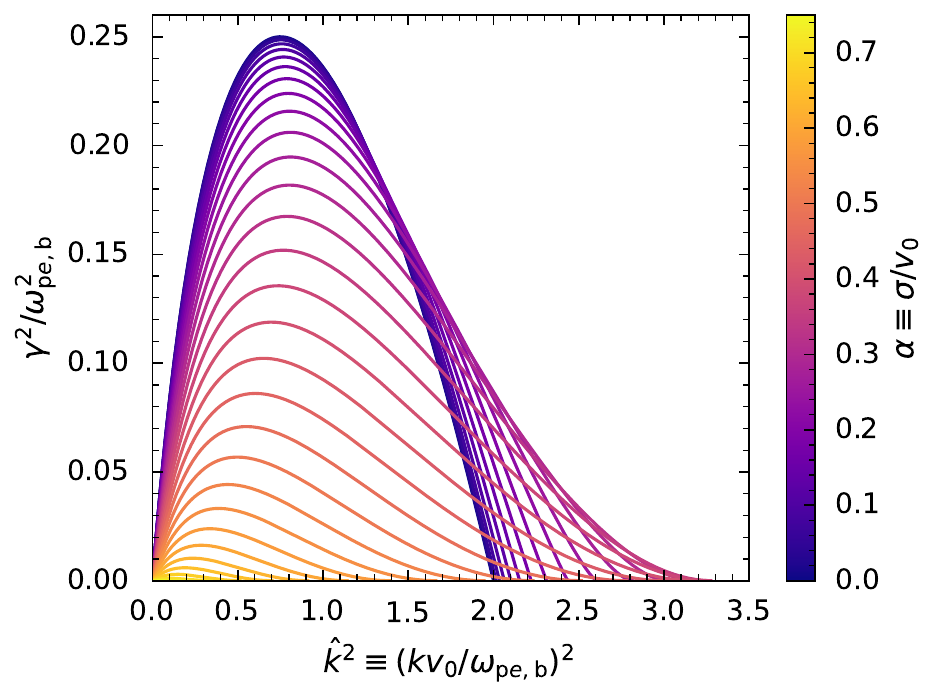}
		\caption{}
		\label{fig:gamma2_vs_k2}
	\end{subfigure}
	\caption{(a) The instability growth rate $\gamma$ divided by the plasma frequency \omegapeb of a beam against $\hat{k}^2$ for two streams with Gaussian velocity distribution functions with standard deviation $\sigma$ centred at $\pm v_0$, for a range of values of $\alpha\equiv \sigma/v_0$ (\cref{eq:gauss_disprel_implicit}). (b) Squares of the same quantities, showing a transition from a linear to a quadratic dependence of $\gamma^2/\omegapeb^2$ on $\hat{k}^2$ near the maximum unstable wavenumber with increasing $\alpha$.}
	\label{fig:gamma_vs_k_and_gamma2_vs_k2}
\end{figure}

\begin{figure}
	\centering
	\begin{subfigure}{0.49\textwidth}
		\includegraphics[width=\textwidth]{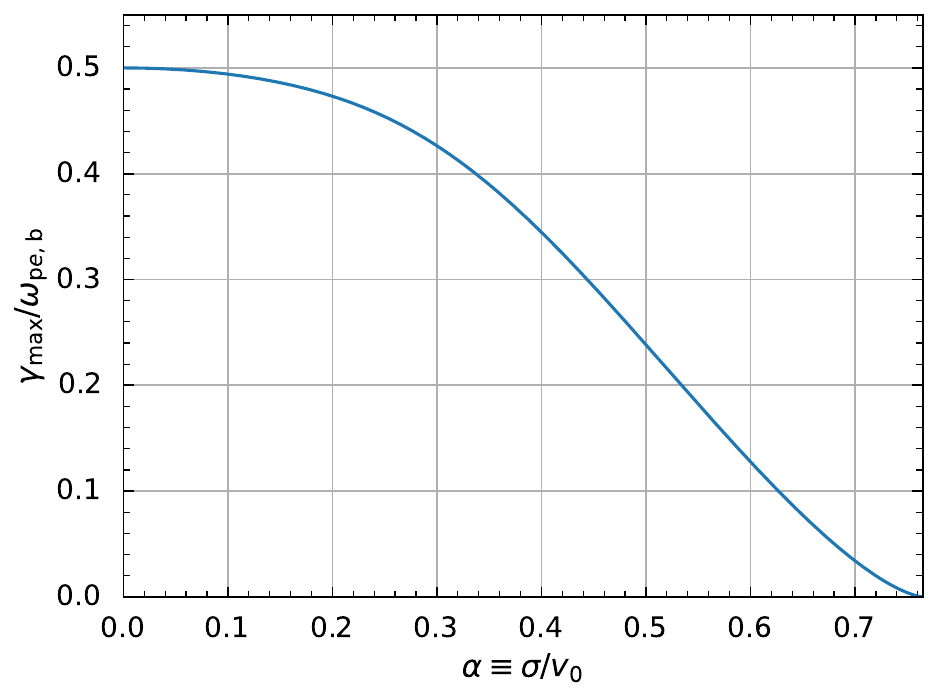}
		\caption{}
		\label{fig:max_growth_rate_vs_alpha}
	\end{subfigure}
	\begin{subfigure}{0.49\textwidth}
		\includegraphics[width=\textwidth]{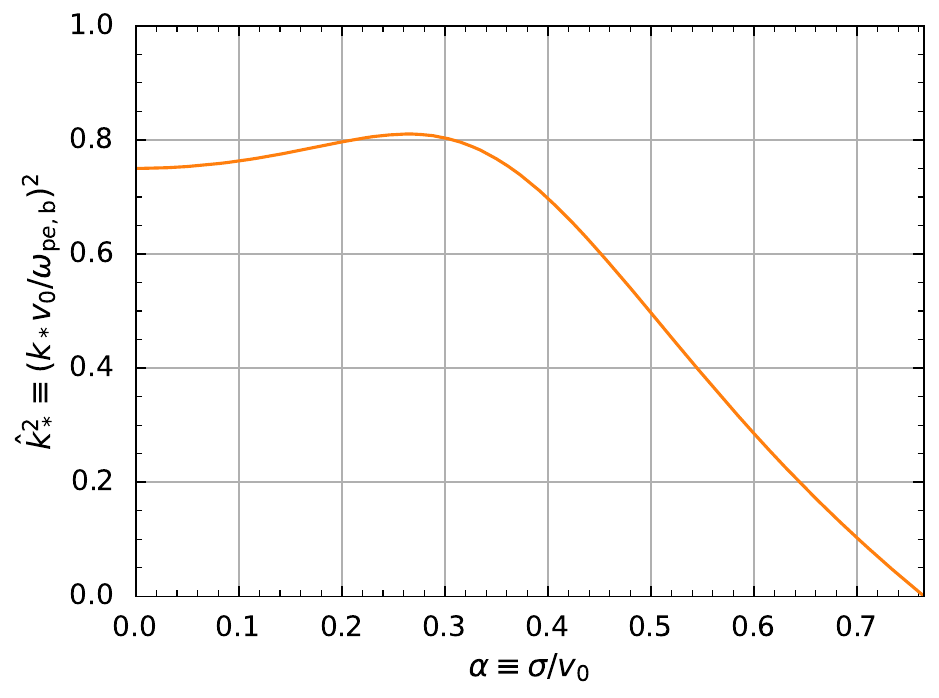}
		\caption{}
		\label{fig:fastest_growing_mode_vs_alpha}
	\end{subfigure}
	\caption{(a) The maximum growth rate divided by the plasma frequency and (b) the wavenumber of the fastest-growing mode of the two-stream instability as a function of the ratio of the standard deviation of the Gaussian distribution to the mean velocity of the beam, $\alpha$.}
\end{figure}

The same asymptotics can be derived for a wider class of distribution functions, without a formal expansion inside the integral, as follows. For two identical beams with distribution function $f_{b}=(n_{b}/\sigma)\tilde{f}((v-v_{b})^{2}/2\sigma^{2}),$ with $\int \tilde{f}(u^{2}/2)\mathrm{d}u=1,$ assuming purely imaginary $\omega=\mathrm{i}\gamma$ ($\gamma>0$) we get
\begin{align}
\nonumber \frac{k^{2}v_{0}^{2}}{\omegapeb^{2}} & = \sum_{v_{b}=\pm v_{0}} v_{0}^{2} \int_{-\infty}^{\infty} \frac{(v_{b}+\sigma u)u\tilde{f}^{\prime}(u^{2}/2) \mathrm{d}u/\sigma}{(v_{b}+\sigma u)^{2}+\gamma^{2}/k^{2}}\\
 & = 2 \int_{-\infty}^{\infty} \frac{(1-\alpha^{2}u^{2} - \uph^{2})}{((1+\alpha u)^{2}+\uph^{2})((1-\alpha u)^{2}+\uph^{2})}(-\tilde{f}^{\prime}(u^{2}/2))u^{2}\mathrm{d}u\ \equiv\ \kappa^{2}(\alpha,\uph^{2})
\label{eq:gauss_disprel_implicit}
\end{align}
with $\alpha=\sigma/v_{0}$ and $\uph=\gamma/(k v_{0}).$ Besides the integration over velocities, the exact expression above is quite similar to \cref{eq:spiky,eq:asympgauss}, and it can be easily shown that the limit for $\alpha\rightarrow0$ is independent of the functional form of $\tilde{f}.$ At fixed $k^{2},$ we can bring partial derivatives under the integration and obtain 
\begin{equation}
\frac{\partial \uph^{2}}{\partial \alpha^{2}} = -\frac{(1/2\alpha)\partial \kappa^{2}/\partial \alpha}{\partial \kappa^{2}/\partial \uph^{2}}\ \sim  3\frac{\uph^{4}-6\uph^{2}+1}{\uph^{4}-2\uph^{2}-3}(\vth^{2}/\sigma^{2}) \ + \alpha^{2}u^{\prime\prime}(\alpha^{2},\uph)\ , \label{eq:asymp_doneright}
\end{equation}
where $\vth^{2}/\sigma^{2}\equiv\int u^{2}\tilde{f}(u^{2}/2)\mathrm{d}u/\int \tilde{f}(u^{2}/2)\mathrm{d}u$ is 1 for Maxwellian beams, and $u^{\prime\prime}\sim\mathcal{O}(1)$ when $\alpha\rightarrow 0.$ The same expansion in \cref{eq:asympgauss} is then recovered by analytical continuation in $\uph^{2}\in\mathbb{C},$ by setting $\hat{v}_{\text{ph}}^{2}=(\omega/k v_{0})^{2}=-\uph^{2}$ in \cref{eq:asymp_doneright}. From the expansion above we can derive the upper and lower plasma modes:
\begin{equation}
  \omega^{2}\ \approx \left\{
    \begin{array}{lll}
      \omegapeb^{2}\left(1+\sqrt{1+6\hat{k}^{2}(1+\alpha^{2})}\right)\  \approx\ \omega_{p}^{2}+3(v_{0}^{2}+\sigma^{2}) k^{2} \ \ \   
      & \mathrm{at} & \ k^{2}\ll \omegapeb^{2}/v_{0}^{2}\\
      \\ 
      \frac{1-\sigma^{2}/v_{0}^{2}}{3}\left(\frac{k^{2}v_{0}^{2}/\omegapeb^{2}}{2} -(1+3\sigma^{2}/v_{0}^{2})\right) \  \  \  
      & \mathrm{at} & \ k^{2}\approx 2\omegapeb^{2}(1+3\sigma^{2}/v_{0}^{2})/v_{0}^{2}
    \end{array}
  \right.
  \label{eq:kappa_integral}
\end{equation}
and, with $\omega=\mathrm{i}\gamma$ ($\gamma>0$), the growth rates:
\begin{equation}
  \gamma^{2}/k^{2}v_{0}^{2}\ \approx \left\{
    \begin{array}{lll}
      (1-6 \sigma^{2}/v_{0}^{2} - 2k^{2}v_{0}^{2}/\omegapeb^{2} )/ (1-3\sigma^{2}/v_{0}^{2})\ \ \   
      & \mathrm{at} & \ k^{2}\ll \omegapeb^{2}/v_{0}^{2}\\
      \\
    \frac{1-\sigma^{2}/v_{0}^{2}}{3}\left(1+3\sigma^{2}/v_{0}^{2}-\frac{k^{2}v_{0}^{2}/\omegapeb^{2}}{2} \right)\ \ \   
      & \mathrm{at} & \ k^{2}\approx 2\omegapeb^{2}(1+3\sigma^{2}/v_{0}^{2})/v_{0}^{2}
    \end{array}
  \right.\ .
\end{equation}
These approximations would predict a null growth rate at $\sigma\approx v_{0}/\sqrt{6},$ beyond the assumed regime $\sigma\ll v_{0}$ of cold beams.
The full behaviour of the growth rate, which reaches zero at $\alpha\approx0.76,$ can be found numerically by scanning values of $0<\uph<1$ 
and consequently $k^2$ from the $\kappa^{2}$ integral in \cref{eq:kappa_integral}. The dependence of the growth rate on $\hat{k}^2$ is shown in \cref{fig:gamma_vs_k_and_gamma2_vs_k2} for a range of values of $\sigma/v_0$, and the relationship between the maximum growth rate and $\sigma/v_0$ is shown in \cref{fig:max_growth_rate_vs_alpha}. The wavenumber of the fastest-growing mode is shown in \cref{fig:fastest_growing_mode_vs_alpha}.


\newpage
The threshold $\sigma\approx v_{0}/\sqrt{6}$ from the linear expansion coincides with the widest wavenumber window of instability ($\hat{k}_{\text{max}}^{2}\approx3$), where the maximum growth rate is attained at $\hat{k}_{\star}^{2}\approx3/4,$ and then $\hat{k}_{\star}^{2}\approx 2(3/4-\sigma/v_{0})$ for $1/\sqrt{6}<\sigma/v_{0}<3/4.$ For $\sigma<0.2v_{0},$ the expression $\gamma^{2}/\omegapeb^{2}\approx (1-5\sigma^{2}/2v_{0}^{2})/4$ is a good approximation to the squared maximum growth rate. The widening of the instability window for small $\alpha$ also explains the behaviour seen in simulations with collisionality~\cite{Hou2024}, where the collisional broadening of cold streams excites instabilities also at wavenumbers that would be stable in the cold limit. At higher $\alpha,$ the growth rate is further reduced by Landau damping, not included here.

\section{Appendix: Hyperparameter optimization}
\label{appendix:hyperparameter_optimisation}

Hyperparameter optimization was performed to identify model and training parameters that led to performant models after fitting to the training data. The MSE on the particle position averaged over all particles, time steps and full rollouts for the validation set was used to determine the performance of a model with a given hyperparameter configuration. Asynchronous successive halving (ASHA)~\cite{asha} was used for scheduling and early stopping of trials, and the tree-structured Parzen estimator (TPE) algorithm~\cite{tpe} was used for hyperparameter configuration selection. These methods are described briefly in the following sections.

The Ray Tune framework~\cite{tune} was used for experiment execution and tracking, and the Optuna~\cite{optuna} implementation of TPE was used for hyperparameter configuration selection. 
200 trials were conducted, each up to a maximum of 2 million gradient updates, which was sufficient for the validation loss to plateau in all viable trials, and took 10 weeks with 12 trials running concurrently, each on one Nvidia A100 GPU. 
The batch size was limited by GPU memory constraints, so it was not optimized and set to 2.
No significant correlation between the minimum rollout MSE and the value of any single hyperparameter was observed.

\subsection{Asynchronous successive halving}

The ASHA algorithm seeks to provide a resource-efficient approach to hyperparameter selection in cases where intermediate performance results are available and are indicators of final performance~\cite{asha}.
A fixed resource budget (for example, computational resource time or training epochs) is defined at the start of the scheduling process, as is a fixed number of brackets.
A reduction factor is chosen, by which the number of models progressing from one bracket to the next is reduced.
The resource budget is distributed equally across the brackets, accounting for the reduction in models from one bracket to the next.

Multiple models with unique hyperparameter configurations are initialized. Models are promoted from one bracket to the next asynchronously, based on their performance relative to other models in the bracket. Once the resource budget is exhausted, the most performant model is selected. Asynchronous execution enables non-blocking promotion of models from one bracket to the next.

The ASHA scheduler was configured to use a reduction factor of 4, one bracket, and train each trial for at least \num{5e5} and at most \num{2e6} gradient updates.

\subsection{Tree-structured Parzen estimator}

The TPE algorithm is a Bayesian optimization method, employing a probabilistic model-based approach to selecting promising hyperparameter configurations by using information on the performance of previous configurations~\cite{tpe}. TPE models two separate probability density functions: $l(\vect{x})$ as the likelihood over hyperparameter trials with performance better than some threshold, and $g(\vect{x})$ as the likelihood over trials with performance worse than the threshold, where $\vect{x}$ are the hyperparameters. The next-most-promising hyperparameter configuration is selected by optimizing expected improvement, which can be shown to be proportional to $l(\vect{x})/g(\vect{x})$.

A Gaussian prior was imposed. So-called \textit{magic clipping}, a heuristic to limit the smallest variances of Gaussians used in the Parzen estimator, was enabled.
In a warm-up phase, the hyperparameter configurations given in \cref{tab:initial_params} were used, as well as additional ones drawn from uniform or logarithmically uniform distributions over the spaces specified in \cref{tab:hyperparameter_search_space}, until 10 trials finished. To calculate the expected improvement, 24 candidate samples were used. 

\subsection{Search space}
\label{search_space}
\setcellgapes{1.5pt} 
\makegapedcells 
\newcolumntype{C}{>{\centering\arraybackslash}X}
\newcolumntype{R}{>{\raggedright}p{0.3\linewidth}}
\begin{table}[ht]
	\caption{Initial sets of configurations for hyperparameter optimization. The values in bold differ from those in Set 1, which are based on the defaults in Ref.~\cite{pytorchGNS}.}
	\label{tab:initial_params}
	\centering
	\begin{tabularx}{\textwidth}{RCCCCCCCC}
		\toprule
		Parameter & Set 1 & Set 2 & Set 3 & Set 4 & Set 5 & Set 6 & Set 7 & Set 8 \\
		\midrule
		Initial learning rate $\eta_0$ & \num{e-4} & \num{e-4} & \num{e-4} & \num{e-4} & \num{e-4} & \num{e-4} & $\mathbf{10^{-6}}$ & \num{e-4} \\
		Learning rate decay factor $d$ & 0.1 & 0.1 & 0.1 & 0.1 & 0.1 & 0.1 & 0.1 & 0.1 \\
		Learning rate drop rate $r$ (millions of steps) & 5 & 5 & 5 & 5 & 5 & 5 & 5 & 5 \\
		Standard deviation of velocity noise in the last step [\unit{\milli\metre\per\second}] & 0.67 & \textbf{0} & 0.67 & \textbf{0} & \textbf{6.7} & \textbf{6.7} & \textbf{6.7} & 0.67 \\
		Standard deviation of electric field noise in the last step [\unit{\newton\per\coulomb}] & \num{e-10} & \textbf{0} & \num{e-10} & \textbf{0} & \num{e-10} & \num{e-10} & \num{e-10} & \num{e-10} \\
		Number of message-passing steps & 10 & 10 & \textbf{12} & \textbf{12} & 10 & \textbf{12} & \textbf{12} & 10 \\
		Number of snapshots in input sequence & 6 & 6 & 6 & 6 & 6 & 6 & 6 & 6 \\
		Connectivity radius [\unit{km}] & 7.5 & 7.5 & 7.5 & 7.5 & 7.5 & 7.5 & 7.5 & \textbf{1.251} \\
		Number of hidden layers in MLPs & 2 & 2 & 2 & 2 & 2 & 2 & 2 & 2 \\
		Number of latent features & 128 & 128 & 128 & 128 & 128 & 128 & 128 & 128 \\
		Number of nodes in each hidden layer of MLPs & 128 & 128 & 128 & 128 & 128 & 128 & 128 & 128 \\
		\bottomrule
	\end{tabularx}
\end{table}

\newcolumntype{P}[1]{>{\centering\arraybackslash}p{#1\linewidth}}
\begin{table}[ht]
    \caption{Hyperparameter search space, the distribution values were drawn from in the warm-up phase, and the parameters of the most performant model, achieving the lowest MSE on the predicted positions in validation set rollouts.}
    \label{tab:hyperparameter_search_space}
    \begin{tabularx}{\textwidth}{RP{0.2}P{0.22}C}
        \toprule
        Hyperparameter & Range & Distribution & Optimal parameters \\
        \midrule
        Initial learning rate $\eta_0$                                                     & \num{e-7} to \num{e-2}                    & Log-uniform & \num[round-mode=figures, round-precision=3, scientific-notation=true]{0.0009714555159814416} \\
        Learning rate decay factor $d$                                                     & \num{e-6} to 10                           & Log-uniform & \num[scientific-notation=true, round-mode=figures, round-precision=2]{0.004378300173832424} \\
        \makecell[cl]{Learning rate drop rate $r$ \\ (number of steps)}                                      & \num{e5} to \num{e9}                      & Log-uniform & \num[scientific-notation=true, round-mode=figures, round-precision=3]{769410.8846882174} \\
        \makecell[cl]{Standard deviation of velocity \\ noise in the last step [\unit{\metre\per\second}]}   & \makecell{0 (5\% chance) or \\ \num{e-6} to \num{2e-2}} & \makecell{Log-uniform \\ (for the non-zero values)} & \num[scientific-notation=true, round-mode=figures, round-precision=3]{1.1441476875355638e-06} \\
        \makecell[cl]{Standard deviation of electric field \\ noise in the last step [\unit{\newton\per\coulomb}]} & \makecell{0 (5\% chance) or \\ \num{e-18} to \num{e-9}} & \makecell{Log-uniform \\ (for the non-zero values)} & \num[scientific-notation=true, round-mode=figures, round-precision=3]{3.027371262125036e-17} \\
        Number of message-passing steps                                                    & 4 to 13                                   & Uniform & 11 \\
        \makecell[cl]{Number of snapshots in input \\ sequence}                                              & 3 to 9                                    & Uniform & 8 \\
        Connectivity radius [\unit{km}]                                                    & 0.625 to 10                               & Uniform & \num[round-mode=figures, round-precision=2]{2.532908790395294} \\
        Number of hidden layers in MLPs                                                    & 1 to 4                                    & Uniform & 1 \\
        Number of latent features                                                          & 20 to 300                                 & Uniform & 209 \\
        \makecell[cl]{Number of nodes in each hidden \\ layer of MLPs}                                       & 20 to 300                                 & Uniform & 185 \\
        \bottomrule
    \end{tabularx}
\end{table}

The initial set of hyperparameters considered, Set 1 in \cref{tab:initial_params}, uses the defaults in Kumar and Vantassel's GNS implementation~\cite{pytorchGNS}, the same connectivity radius relative to the domain size used in most physical domains by Sanchez-Gonzalez et al.~\cite{deepmind}, and a standard deviation of the electric field noise equal to that of the velocity noise relative to the maximum values in the training dataset. It is the same configuration used for Models A and B, which were trained without field nodes, but with the addition of electric field noise.

The rest of the initial hyperparameter configurations modify some of the most important parameters identified by Sanchez-Gonzalez et al.: the noise level, which was set to 0 to compare results with a model trained without added noise, as well as to a larger value than the default, the number of message-passing steps, which was increased to 12, the maximum value which did not result in exceeding the memory capacity of a V100 GPU, and the connectivity radius, which was set to just above the grid resolution, to ensure that neighbouring field nodes are connected. 

Set 8 also included a reduced initial learning rate, to try to reduce large fluctuations of the loss which were observed during training of models A and B (see \cref{fig:validation_loss}). For other trials, the search space was expanded to also allow for a larger initial, but more rapidly decaying learning rate. As seen in \cref{fig:validation_loss}, such a learning rate schedule indeed resulted in a more stable training for model C.

Adding noise to input velocities and electric fields was intended to help the model predict correct acceleration even from imperfect inputs and thus mitigate the accumulation of error from its predictions over long rollouts. However, the optimum values of the standard deviation of the noise were set at negligible values near the bottom of the explored range, indicating that the input corruption was not helpful. 

When it comes to choosing the number of previous snapshots used to define input features, we might expect that adding velocities and electric field values from more snapshots should help the model better predict the next one. 
However, snapshots further in the past will be less useful, and adding more might result in overfitting and hinder the simulator's performance.

In principle, increasing the connectivity radius and adding more message-passing steps allows each node to be influenced by more distant nodes, which could be beneficial for the model to learn the dynamics of the system, and especially as longer time steps are used. However, the maximum number of message-passing steps and connectivity radius are limited by available GPU memory, and making these parameters too large may result in issues such as oversquashing~\cite{oversquashing}, where information flowing from distant nodes is distorted, and oversmoothing~\cite{oversmoothing}, which means node features become more similar with more message-passing steps.

Other important physical considerations are Debye shielding, which means that the electric field of a charge is greatly screened by oppositely charged particles at distances beyond the Debye length, $\lambda_\text{D}=\vth/\omega_\text{p}$, so connections to far away particles may be irrelevant, and the grid resolution --- nodes should receive information from neighbouring grid cells. 
The minimum connectivity radius considered was half the grid resolution, which would mean that each particle node is connected to one field node, or two if it is at the centre of a cell, and would only gain information about the field at other grid points through repeated message passing. The upper bound on the connectivity radius search space was set to slightly higher than the maximum Debye length occurring in the dataset. 

The connectivity radius of the best-performing model was twice the grid resolution. 
The number of message-passing steps and input sequence length were close to the upper bounds of the considered search space, suggesting further increasing them could improve performance.  However, it would also increase simulation run time and require more GPU memory, and performance improvement is expected to diminish.

\printbibliography{}

@article{plasmaGNS,
title = {Learning the dynamics of a one-dimensional plasma model with graph neural networks},
author = {Diogo D Carvalho and Diogo R Ferreira and Luís O Silva},
journal = {Machine Learning: Science and Technology},
shortjournal = {Mach. Learn.: Sci. Technol.},
year = {2024},
month = {5},
volume = {5},
number = {2},
pages = {025048},
doi = {10.1088/2632-2153/ad4ba6},
url = {https://dx.doi.org/10.1088/2632-2153/ad4ba6},
publisher = {IOP Publishing}
}

@ARTICLE{Hou2024,
       author = {{Hou}, Y.~W. and {Yu}, M.~Y. and {Wang}, J.~F. and {Liu}, C.~Y. and {Chen}, M.~X. and {Wu}, B.},
        title = {Suppression and excitation by collisions of two-stream and bump-on-tail instabilities},
      journal = {Physics of Plasmas},
      shortjournal={Phys. Plasmas},
         year = 2024,
        month = dec,
       volume = {31},
       number = {12},
          eid = {122103},
        pages = {122103},
          doi = {10.1063/5.0238753},
       adsurl = {https://ui.adsabs.harvard.edu/abs/2024PhPl...31l2103H}
}

@BOOK{Krall,
       author = {{Krall}, Nicholas A. and {Trivelpiece}, A.~W.},
        title = {Principles of plasma physics},
        publisher = {McGraw-Hill},
         year = 1973,
         isbn = {0-07-035346-8}
}

@ARTICLE{Penrose1960,
       author = {Penrose, Oliver},
        title = {Electrostatic Instabilities of a Uniform Non-{Maxwellian} Plasma},
      journal = {Physics of Fluids},
      shortjournal = {Phys. Fluids},
         year = 1960,
        month = mar,
       volume = {3},
       number = {2},
        pages = {258-265},
          doi = {10.1063/1.1706024},
       adsurl = {https://ui.adsabs.harvard.edu/abs/1960PhFl....3..258P}
}

@InProceedings{deepmind,
  title = 	 {Learning to Simulate Complex Physics with Graph Networks},
  author =       {Sanchez-Gonzalez, Alvaro and Godwin, Jonathan and Pfaff, Tobias and Ying, Rex and Leskovec, Jure and Battaglia, Peter},
  booktitle = 	 {Proceedings of the 37th International Conference on Machine Learning},
  pages = 	 {8459--8468},
  year = 	 {2020},
  editor = 	 {III, Hal Daumé and Singh, Aarti},
  volume = 	 {119},
  series = 	 {Proceedings of Machine Learning Research},
  month = 	 {7},
  publisher =    {PMLR},
  url = 	 {https://proceedings.mlr.press/v119/sanchez-gonzalez20a.html}
}

@article{pytorchGNS, 
  doi = {10.21105/joss.05025},
  year = {2023}, 
  publisher = {The Open Journal}, 
  volume = {8}, 
  number = {88}, 
  pages = {5025}, 
  author = {Krishna Kumar and Joseph Vantassel}, 
  title = {{{GNS}}: A generalizable Graph Neural Network-based simulator for particulate and fluid modeling}, 
  journal = {Journal of Open Source Software},
  shortjournal = {J. Open Source Softw.},
}

@article{epoch,
doi = {10.1088/0741-3335/57/11/113001},
year = {2015},
month = {9},
publisher = {IOP Publishing},
volume = {57},
number = {11},
pages = {113001},
author = {T D Arber and K Bennett and C S Brady and A Lawrence-Douglas and M G Ramsay and N J Sircombe and P Gillies and R G Evans and H Schmitz and A R Bell and C P Ridgers},
title = {Contemporary particle-in-cell approach to laser-plasma modelling},
journal = {Plasma Physics and Controlled Fusion},
shortjournal = {Plasma Phys. Control. Fusion}
}

@inproceedings{inverseDesign,
 author = {Allen, Kelsey and Lopez-Guevara, Tatiana and Stachenfeld, Kimberly L and Sanchez Gonzalez, Alvaro and Battaglia, Peter and Hamrick, Jessica B and Pfaff, Tobias},
 booktitle = {Advances in Neural Information Processing Systems},
 editor = {S. Koyejo and S. Mohamed and A. Agarwal and D. Belgrave and K. Cho and A. Oh},
 pages = {13759--13774},
 publisher = {Curran Associates, Inc.},
 title = {Inverse Design for Fluid--Structure Interactions using Graph Network Simulators},
 url = {https://proceedings.neurips.cc/paper_files/paper/2022/hash/59593615e358d52295578e0d8e94ec4a-Abstract-Conference.html},
 volume = {35},
 year = {2022}
}

@article{diffusionGenerativeInverseDesign,
      title={Diffusion Generative Inverse Design}, 
      author={Marin Vlastelica and Tatiana López-Guevara and Kelsey Allen and Peter Battaglia and Arnaud Doucet and Kimberley Stachenfeld},
      year={2023},
      eprint={2309.02040},
      archivePrefix={arXiv},
      primaryClass={cs.LG}
}

@inproceedings{KumarChoi2023,
author = {Kumar, Krishna and Choi, Yonjin},
title = {Accelerating Particle and Fluid Simulations with Differentiable Graph Networks for Solving Forward and Inverse Problems},
year = {2023},
isbn = {979-8-4007-0785-8},
publisher = {Association for Computing Machinery},
address = {New York, NY, USA},
url = {https://doi.org/10.1145/3624062.3626082},
doi = {10.1145/3624062.3626082},
booktitle = {Proceedings of the SC '23 Workshops of The International Conference on High Performance Computing, Network, Storage, and Analysis},
pages = {60–65},
numpages = {6},
location = {Denver, CO, USA},
series = {SC-W '23}
}

@InProceedings{GNcontrol,
  title = 	 {Graph Networks as Learnable Physics Engines for Inference and Control},
  author =       {Sanchez-Gonzalez, Alvaro and Heess, Nicolas and Springenberg, Jost Tobias and Merel, Josh and Riedmiller, Martin and Hadsell, Raia and Battaglia, Peter},
  booktitle = 	 {Proceedings of the 35th International Conference on Machine Learning},
  pages = 	 {4470--4479},
  year = 	 {2018},
  editor = 	 {Dy, Jennifer and Krause, Andreas},
  volume = 	 {80},
  series = 	 {Proceedings of Machine Learning Research},
  month = 	 {7},
  publisher =    {PMLR},
  url = 	 {https://proceedings.mlr.press/v80/sanchez-gonzalez18a.html}
}

@ARTICLE{gnn_journal,
  author={Scarselli, Franco and Gori, Marco and Tsoi, Ah Chung and Hagenbuchner, Markus and Monfardini, Gabriele},
  journal={IEEE Transactions on Neural Networks}, 
  shortjournal={IEEE Trans. Neural Netw.},
  title={The Graph Neural Network Model}, 
  year={2009},
  volume={20},
  number={1},
  pages={61-80},
  doi={10.1109/TNN.2008.2005605}
}

@book{birdsallPlasmaPhysicsComputer2017,
  title = {Plasma Physics via Computer Simulation},
  author = {Birdsall, C. K. and Langdon, A. B.},
  year = {2017},
  month = jan,
  publisher = {CRC Press},
  address = {Boca Raton},
  doi = {10.1201/9781315275048},
  isbn = {978-1-315-27504-8}
}

@article{dijkPlasmaModellingNumerical2009,
  title = {Plasma modelling and numerical simulation},
  author = {van Dijk, J and Kroesen, G M W and Bogaerts, A},
  year = {2009},
  month = {9},
  journal = {Journal of Physics D: Applied Physics},
  shortjournal = {J. Phys. D: Appl. Phys.},
  volume = {42},
  number = {19},
  pages = {190301},  
  doi = {10.1088/0022-3727/42/19/190301},
  url = {https://dx.doi.org/10.1088/0022-3727/42/19/190301}
}

@inproceedings{pytorch,
author = {Ansel, Jason and Yang, Edward and He, Horace and Gimelshein, Natalia and Jain, Animesh and Voznesensky, Michael and Bao, Bin and Bell, Peter and Berard, David and Burovski, Evgeni and Chauhan, Geeta and Chourdia, Anjali and Constable, Will and Desmaison, Alban and DeVito, Zachary and Ellison, Elias and Feng, Will and Gong, Jiong and Gschwind, Michael and Hirsh, Brian and Huang, Sherlock and Kalambarkar, Kshiteej and Kirsch, Laurent and Lazos, Michael and Lezcano, Mario and Liang, Yanbo and Liang, Jason and Lu, Yinghai and Luk, CK and Maher, Bert and Pan, Yunjie and Puhrsch, Christian and Reso, Matthias and Saroufim, Mark and Siraichi, Marcos Yukio and Suk, Helen and Suo, Michael and Tillet, Phil and Wang, Eikan and Wang, Xiaodong and Wen, William and Zhang, Shunting and Zhao, Xu and Zhou, Keren and Zou, Richard and Mathews, Ajit and Chanan, Gregory and Wu, Peng and Chintala, Soumith},
booktitle = {29th ACM International Conference on Architectural Support for Programming Languages and Operating Systems, Volume 2 (ASPLOS '24)},
doi = {10.1145/3620665.3640366},
month = apr,
publisher = {Association for Computing Machinery},
title = {{PyTorch} 2: Faster Machine Learning Through Dynamic {Python} Bytecode Transformation and Graph Compilation},
url = {https://pytorch.org/assets/pytorch2-2.pdf},
year = {2024}
}

@inproceedings{pytorch_geometric,
  title={Fast Graph Representation Learning with {PyTorch Geometric}},
  author={Fey, Matthias and Lenssen, Jan E.},
  booktitle={ICLR Workshop on Representation Learning on Graphs and Manifolds},
  year={2019},
  url={https://rlgm.github.io/papers/2.pdf}
}

@article{DEROUILLAT2018351,
title = {Smilei : {A} collaborative, open-source, multi-purpose particle-in-cell code for plasma simulation},
journal = {Computer Physics Communications},
shortjournal = {Comput. Phys. Comm.},
volume = {222},
pages = {351-373},
year = {2018},
issn = {0010-4655},
doi = {https://doi.org/10.1016/j.cpc.2017.09.024},
url = {https://www.sciencedirect.com/science/article/pii/S0010465517303314},
author = {J. Derouillat and A. Beck and F. Pérez and T. Vinci and M. Chiaramello and A. Grassi and M. Flé and G. Bouchard and I. Plotnikov and N. Aunai and J. Dargent and C. Riconda and M. Grech}
}

@article{LIFSCHITZ20091803,
title = {Particle-in-Cell modelling of laser–plasma interaction using {Fourier} decomposition},
journal = {Journal of Computational Physics},
shortjournal = {J. Computat. Phys.},
volume = {228},
number = {5},
pages = {1803-1814},
year = {2009},
issn = {0021-9991},
doi = {https://doi.org/10.1016/j.jcp.2008.11.017},
url = {https://www.sciencedirect.com/science/article/pii/S0021999108005950},
author = {A.F. Lifschitz and X. Davoine and E. Lefebvre and J. Faure and C. Rechatin and V. Malka}
}

@article{GODFREY2014689,
title = {Numerical stability analysis of the pseudo-spectral analytical time-domain {PIC} algorithm},
journal = {Journal of Computational Physics},
shortjournal = {J. Computat. Phys.},
volume = {258},
pages = {689-704},
year = {2014},
issn = {0021-9991},
doi = {https://doi.org/10.1016/j.jcp.2013.10.053},
url = {https://www.sciencedirect.com/science/article/pii/S0021999113007298},
author = {Brendan B. Godfrey and Jean-Luc Vay and Irving Haber}
}

@article{PhysRevLett.98.130405,
  title = {Noninvariance of Space- and Time-Scale Ranges under a {Lorentz} Transformation and the Implications for the Study of Relativistic Interactions},
  author = {Vay, J.-L.},
  journal = {Phys. Rev. Lett.},
  volume = {98},
  issue = {13},
  pages = {130405},
  numpages = {4},
  year = {2007},
  month = {3},
  publisher = {American Physical Society},
  doi = {10.1103/PhysRevLett.98.130405},
  url = {https://link.aps.org/doi/10.1103/PhysRevLett.98.130405}
}

@article{martins2010exploring,
  title={Exploring laser-wakefield-accelerator regimes for near-term lasers using particle-in-cell simulation in {Lorentz}-boosted frames},
  author={Martins, Samuel F. and Fonseca, R. A. and Lu, Wei and Mori, Warren B. and Silva, L. O.},
  journal={Nature Physics},
  shortjournal={Nature Phys.},
  volume={6},
  number={4},
  pages={311--316},
  year={2010},
  month={4},
  publisher={Nature Publishing Group UK London},
  issn = {1745-2481},
  doi = {10.1038/nphys1538}
}

@inproceedings{10.1145/2503210.2504564,
author = {Bussmann, M. and Burau, H. and Cowan, T. E. and Debus, A. and Huebl, A. and Juckeland, G. and Kluge, T. and Nagel, W. E. and Pausch, R. and Schmitt, F. and Schramm, U. and Schuchart, J. and Widera, R.},
title = {Radiative signatures of the relativistic {Kelvin}--{Helmholtz} instability},
year = {2013},
isbn = {978-1-4503-2378-9},
publisher = {Association for Computing Machinery},
address = {New York, NY, USA},
url = {https://doi.org/10.1145/2503210.2504564},
doi = {10.1145/2503210.2504564},
booktitle = {Proceedings of the International Conference on High Performance Computing, Networking, Storage and Analysis},
articleno = {5},
numpages = {12},
location = {Denver, Colorado},
series = {SC '13}
}

@INPROCEEDINGS {10046112,
author = { Fedeli, Luca and Huebl, Axel and Boillod-Cerneux, France and Clark, Thomas and Gott, Kevin and Hillairet, Conrad and Jaure, Stephan and Leblanc, Adrien and Lehe, Remi and Myers, Andrew and Piechurski, Christelle and Sato, Mitsuhisa and Zaim, Neil and Zhang, Weiqun and Vay, Jean-Luc and Vincenti, Henri },
booktitle = { SC22: International Conference for High Performance Computing, Networking, Storage and Analysis },
title = {Pushing the Frontier in the Design of Laser-Based Electron Accelerators with Groundbreaking Mesh-Refined Particle-In-Cell Simulations on Exascale-Class Supercomputers},
year = {2022},
volume = {},
ISSN = {},
pages = {1-12},
doi = {10.1109/SC41404.2022.00008},
url = {https://doi.ieeecomputersociety.org/10.1109/SC41404.2022.00008},
publisher = {IEEE Computer Society},
address = {Los Alamitos, CA, USA},
month =Nov}

@article{nerush2011laser,
  title = {Laser Field Absorption in Self-Generated Electron--Positron Pair Plasma},
  author = {Nerush, E. N. and Kostyukov, I. Yu. and Fedotov, A. M. and Narozhny, N. B. and Elkina, N. V. and Ruhl, H.},
  journal={Physical Review Letters},
  shortjournal = {Phys. Rev. Lett.},
  volume = {106},
  issue = {3},
  pages = {035001},
  numpages = {4},
  year = {2011},
  month = {1},
  publisher = {American Physical Society},
  doi = {10.1103/PhysRevLett.106.035001},
  url = {https://link.aps.org/doi/10.1103/PhysRevLett.106.035001}
}

@article{ridgers2012dense,
  title = {Dense Electron--Positron Plasmas and Ultraintense $\gamma$ rays from Laser-Irradiated Solids},
  author = {Ridgers, C. P. and Brady, Christopher S. and Duclous, R. and Kirk, J. G. and Bennett, K. and Arber, T. D. and Robinson, A. P. L. and Bell, A. R.},
  journal={Physical Review Letters},
  shortjournal = {Phys. Rev. Lett.},
  volume = {108},
  issue = {16},
  pages = {165006},
  numpages = {5},
  year = {2012},
  month = {4},
  publisher = {American Physical Society},
  doi = {10.1103/PhysRevLett.108.165006},
  url = {https://link.aps.org/doi/10.1103/PhysRevLett.108.165006}
}

@article{HORNIK1989359,
title = {Multilayer feedforward networks are universal approximators},
journal = {Neural Networks},
shortjournal = {Neural Netw.},
volume = {2},
number = {5},
pages = {359-366},
year = {1989},
issn = {0893-6080},
doi = {https://doi.org/10.1016/0893-6080(89)90020-8},
url = {https://www.sciencedirect.com/science/article/pii/0893608089900208},
author = {Kurt Hornik and Maxwell Stinchcombe and Halbert White}
}

@book{bishop1995neural,
    title = {Neural Networks for Pattern Recognition},
    author = {Bishop, Christopher M},
    publisher = {Oxford University Press},
    year = {1995},
    month = {11},
    isbn = {978-0-19-853849-3},
    doi = {10.1093/oso/9780198538493.001.0001},
    url = {https://doi.org/10.1093/oso/9780198538493.001.0001},
}

@article{rumelhart1986learning,
  title = {Learning representations by back-propagating errors},
  author = {Rumelhart, David E. and Hinton, Geoffrey E. and Williams, Ronald J.},
  year = {1986},
  month = {10},
  journal = {Nature},
  volume = {323},
  number = {6088},
  pages = {533--536},
  issn = {1476-4687},
  doi = {10.1038/323533a0}
}

@inproceedings{grzeszczuk1998neuroanimator,
  title = {{{NeuroAnimator}}: fast neural network emulation and control of physics-based models},
  author = {Grzeszczuk, Radek and Terzopoulos, Demetri and Hinton, Geoffrey},
  year = {1998},
  isbn = {0897919998},
  publisher = {Association for Computing Machinery},
  address = {New York, NY, USA},
  url = {https://doi.org/10.1145/280814.280816},
  doi = {10.1145/280814.280816},
  booktitle = {Proceedings of the 25th Annual Conference on Computer Graphics and Interactive Techniques},
  pages = {9–20},
  numpages = {12},
  series = {SIGGRAPH '98}
}

@INPROCEEDINGS{1555942,
  author={Gori, M. and Monfardini, G. and Scarselli, F.},
  booktitle={Proceedings. 2005 IEEE International Joint Conference on Neural Networks, vol. 2}, 
  title={A new model for learning in graph domains}, 
  year={2005},
  number={},
  pages={729-734},
  doi={10.1109/IJCNN.2005.1555942}
}

@article{Battaglia2018RelationalIB,
      title={Relational inductive biases, deep learning, and graph networks}, 
      author={Peter W. Battaglia and Jessica B. Hamrick and Victor Bapst and Alvaro Sanchez-Gonzalez and Vinicius Zambaldi and Mateusz Malinowski and Andrea Tacchetti and David Raposo and Adam Santoro and Ryan Faulkner and Caglar Gulcehre and Francis Song and Andrew Ballard and Justin Gilmer and George Dahl and Ashish Vaswani and Kelsey Allen and Charles Nash and Victoria Langston and Chris Dyer and Nicolas Heess and Daan Wierstra and Pushmeet Kohli and Matt Botvinick and Oriol Vinyals and Yujia Li and Razvan Pascanu},
      year={2018},
      eprint={1806.01261},
      archivePrefix={arXiv},
      primaryClass={cs.LG},
      url={https://arxiv.org/abs/1806.01261},
}

@article{Chang2016ACO,
      title={A Compositional Object-Based Approach to Learning Physical Dynamics}, 
      author={Michael B. Chang and Tomer Ullman and Antonio Torralba and Joshua B. Tenenbaum},
      year={2017},
      eprint={1612.00341},
      archivePrefix={arXiv},
      primaryClass={cs.AI},
      url={https://arxiv.org/abs/1612.00341}, 
}

@inproceedings{Battaglia2016InteractionNF,
 author = {Battaglia, Peter and Pascanu, Razvan and Lai, Matthew and Jimenez Rezende, Danilo and Kavukcuoglu, Koray},
 booktitle = {Advances in Neural Information Processing Systems},
 editor = {D. Lee and M. Sugiyama and U. Luxburg and I. Guyon and R. Garnett},
 pages = {},
 publisher = {Curran Associates, Inc.},
 title = {Interaction Networks for Learning about Objects, Relations and Physics},
 url = {https://proceedings.neurips.cc/paper_files/paper/2016/hash/3147da8ab4a0437c15ef51a5cc7f2dc4-Abstract.html},
 volume = {29},
 year = {2016}
}

@article{Tajima2020WakefieldA,
  title = {Wakefield acceleration},
  author = {Toshiki Tajima and X. Q. Yan and Toshikazu Ebisuzaki},
  year = {2020},
  month = may,
  journal = {Reviews of Modern Plasma Physics},
  shortjournal = {Rev. Mod. Plasma Phys.},
  volume = {4},
  number = {1},
  pages = {7},
  issn = {2367-3192},
  doi = {10.1007/s41614-020-0043-z}
}

@article{Macchi2013IonAB,
  title = {Ion acceleration by superintense laser--plasma interaction},
  author = {Andrea Macchi and Marco Borghesi and Matteo Passoni},
  journal = {Reviews of Modern Physics},
  shortjournal = {Rev. Mod. Phys.},
  volume = {85},
  issue = {2},
  pages = {751--793},
  numpages = {0},
  year = {2013},
  month = {5},
  publisher = {American Physical Society},
  doi = {10.1103/RevModPhys.85.751},
  url = {https://link.aps.org/doi/10.1103/RevModPhys.85.751}
}

@article{Lehmann2016TransientPP,
  title = {Transient Plasma Photonic Crystals for High-Power Lasers},
  author = {Lehmann, G. and Spatschek, K. H.},
  journal = {Phys. Rev. Lett.},
  volume = {116},
  issue = {22},
  pages = {225002},
  numpages = {5},
  year = {2016},
  month = {6},
  publisher = {American Physical Society},
  doi = {10.1103/PhysRevLett.116.225002},
  url = {https://link.aps.org/doi/10.1103/PhysRevLett.116.225002}
}

@book{Bchner2003SpacePS,
  title={Space Plasma Simulation},
  author={J{\"o}rg B{\"u}chner and Christian T. Dum and Manfred Scholer},
  publisher = {Springer Berlin},
  year={2003},
  isbn = {978-3-540-36530-3},
  doi = {10.1007/3-540-36530-3},
  url={https://doi.org/10.1007/3-540-36530-3}
}

@article{Birn2012ParticleAI,
  title = {Particle Acceleration in the Magnetotail and Aurora},
  author = {Birn, J. and Artemyev, A. V. and Baker, D. N. and Echim, M. and Hoshino, M. and Zelenyi, L. M.},
  year = {2012},
  month = nov,
  journal = {Space Science Reviews},
  shortjournal = {Space Sci. Rev.},
  volume = {173},
  number = {1},
  pages = {49--102},
  issn = {1572-9672},
  doi = {10.1007/s11214-012-9874-4}
}

@article{Dimits2000ComparisonsAP,
    title = {Comparisons and physics basis of tokamak transport models and turbulence simulations},
    author={Andris M. Dimits and Glenn Bateman and Michael Beer and Bruce I. Cohen and William D. Dorland and Gregory W. Hammett and Charlson C. Kim and J. E. Kinsey and M. T. Kotschenreuther and Arnold Kritz and Lang L. Lao and John Mandrekas and William McCay Nevins and Scott E. Parker and Aaron John Redd and Dan E. Shumaker and Richard D. Sydora and Janet L. Weiland},
    journal = {Physics of Plasmas},
    shortjournal = {Phys. Plasmas},
    volume = {7},
    number = {3},
    pages = {969-983},
    year = {2000},
    month = {03},
    issn = {1070-664X},
    doi = {10.1063/1.873896},
    url = {https://doi.org/10.1063/1.873896},
    eprint = {https://pubs.aip.org/aip/pop/article-pdf/7/3/969/19007104/969\_1\_online.pdf},
}

@book{Colonna2017PlasmaMM,
author = {Colonna, Gianpiero and D'Angola, Antonio},
title = {Plasma Modeling: Methods and Applications},
publisher = {IOP Publishing},
year = {2016},
series = {2053-2563},
isbn = {978-0-7503-1200-4},
url = {https://dx.doi.org/10.1088/978-0-7503-1200-4},
doi = {10.1088/978-0-7503-1200-4}
}

@article{Courant2015OnTP,
  author={Courant, R. and Friedrichs, K. and Lewy, H.},
  journal={IBM Journal of Research and Development},
  shortjournal={IBM J. Res. Dev.},
  title={On the Partial Difference Equations of Mathematical Physics}, 
  year={1967},
  volume={11},
  number={2},
  pages={215-234},
  doi={10.1147/rd.112.0215}
}

@article{Nicolini2019ModelOR,
  title={Model Order Reduction of Electromagnetic Particle-in-Cell Kinetic Plasma Simulations via Proper Orthogonal Decomposition},
  author={Julio L. Nicolini and Dong-Yeop Na and Fernando L. Teixeira},
  journal={IEEE Transactions on Plasma Science},
  shortjournal={IEEE Trans. Plasma Sci.},
  year={2019},
  volume={47},
  number={12},
  pages={5239-5250},
  doi={10.1109/TPS.2019.2950377}}

@article{Nayak2023AcceleratingPK,
  title = {Accelerating particle-in-cell kinetic plasma simulations via reduced-order modeling of space-charge dynamics using dynamic mode decomposition},
  author = {Nayak, Indranil and Teixeira, Fernando L. and Na, Dong-Yeop and Kumar, Mrinal and Omelchenko, Yuri A.},
  journal={Physical review. E},
  shortjournal = {Phys. Rev. E},
  volume = {109},
  issue = {6},
  pages = {065307},
  numpages = {15},
  year = {2024},
  month = {6},
  publisher = {American Physical Society},
  doi = {10.1103/PhysRevE.109.065307},
  url = {https://link.aps.org/doi/10.1103/PhysRevE.109.065307}
}

@article{Hesthaven2022AdaptiveSM,
  title={Adaptive symplectic model order reduction of parametric particle-based {Vlasov--Poisson} equation},
  author={Jan S. Hesthaven and Cecilia Pagliantini and Nicol{\`o} Ripamonti},
  journal={Mathematics of Computation},
  shortjournal={Math. Comput.},
  year={2024},
  volume={93},
  pages={1153-1202},
  doi={10.1090/mcom/3885},
  url={https://doi.org/10.1090/mcom/3885}
}

@INPROCEEDINGS{Aguilar2021ADL,
  title={A Deep Learning-Based Particle-in-Cell Method for Plasma Simulations},
  author={Xavier Aguilar and Stefano Markidis},
  booktitle={2021 IEEE International Conference on Cluster Computing},
  year={2021},
  pages={692-697},
  doi={10.1109/Cluster48925.2021.00103}
}

@article{Kube2021MachineLA,
      title={Machine learning accelerated particle-in-cell plasma simulations}, 
      author={R. Kube and R. M. Churchill and B. Sturdevant},
      year={2021},
      eprint={2110.12444},
      archivePrefix={arXiv},
      primaryClass={physics.plasm-ph},
      url={https://arxiv.org/abs/2110.12444}, 
}

@article{Badiali2022MachinelearningbasedMI,
  title={Machine-learning-based models in particle-in-cell codes for advanced physics extensions},
  author={Chiara Badiali and Pablo J. Bilbao and F{\'a}bio Cruz and Lu{\'i}s O. Silva},
  journal={Journal of Plasma Physics},
  shortjournal={J. Plasma Phys.},
  year={2022},
  volume={88},
  number={6},
  pages={895880602},
  DOI={10.1017/S0022377822001180}
}

@article{Amaro2024NeuralNS,
      title={Neural network sampling of {Bethe--Heitler} process in particle-in-cell codes}, 
      author={Óscar Amaro and Chiara Badiali and Bertrand Martinez},
      year={2024},
      eprint={2406.02491},
      archivePrefix={arXiv},
      primaryClass={physics.comp-ph},
      url={https://arxiv.org/abs/2406.02491}, 
}

@article{Djordjevi2021ModelingLI,
  title={Modeling laser-driven ion acceleration with deep learning},
  author={B. Z. Djordjevi{\'c} and Andreas J. Kemp and J. Kim and Raspberry A. Simpson and Scott C. Wilks and Tammy Ma and D. A. Mariscal},
  journal={Physics of Plasmas},
  shortjournal={Phys. Plasmas},
  year={2021},
  month = {04},
  volume={28},
  number={4},
  pages={043105},
  issn = {1070-664X},
  doi = {10.1063/5.0045449},
  url = {https://doi.org/10.1063/5.0045449},
  eprint = {https://pubs.aip.org/aip/pop/article-pdf/doi/10.1063/5.0045449/16007794/043105\_1\_online.pdf},
}

@article{Liu2024DeepLA,
  title={Deep learning approaches for modeling laser-driven proton beams via phase-stable acceleration},
  author={Yao-Li Liu and Yen-Chen Chen and Chun-Sung Jao and Mao-Syun Wong and Chun-Han Huang and Han-Wei Chen and Shogo Isayama and Yasuhiro Kuramitsu},
  journal={Physics of Plasmas},
  shortjournal={Phys. Plasmas},
  year={2024},
  month = {01},
  volume = {31},
  number = {1},
  pages = {013106},
  issn = {1070-664X},
  doi = {10.1063/5.0178238},
  url = {https://doi.org/10.1063/5.0178238},
  eprint = {https://pubs.aip.org/aip/pop/article-pdf/doi/10.1063/5.0178238/18847907/013106\_1\_5.0178238.pdf},
}

@article{Schmitz2023ModelingOA,
  title={Modeling of a Liquid Leaf Target {TNSA} Experiment Using Particle-In-Cell Simulations and Deep Learning},
  author={B. Schmitz and Daniel Kreuter and Oliver Boine-Frankenheim and Daniele Margarone},
  journal={Laser and Particle Beams},
  shortjournal={Laser Part. Beams},
  volume={2023},
  year={2023},
  pages={e3},
  DOI={10.1155/2023/2868112}
}

@article{Desai2023ApplyingMM,
  title={Applying Machine-Learning Methods to Laser Acceleration of Protons: Lessons Learned From Synthetic Data},
  author={Ronak Desai and Thomas Zhang and John J. Felice and Ricky Oropeza and Joseph R Smith and Alona Kryshchenko and Chris Orban and Michael L. Dexter and Anil K. Patnaik},
  journal={Contributions to Plasma Physics},
  shortjournal={Contrib. Plasma Phys.},
  year={2024},
  pages = {e202400080},
  doi = {https://doi.org/10.1002/ctpp.202400080},
  url = {https://onlinelibrary.wiley.com/doi/abs/10.1002/ctpp.202400080},
  eprint = {https://onlinelibrary.wiley.com/doi/pdf/10.1002/ctpp.202400080}
}

@inproceedings{Sandberg2024SynthesizingPS,
  title = {Synthesizing Particle-In-Cell Simulations through Learning and {GPU} Computing for Hybrid Particle Accelerator Beamlines},
  author = {R. T. Sandberg and R{\'e}mi Lehe and Chad Mitchell and Marco Garten and Andrew Myers and Ji Qiang and Jean-Luc Vay and Axel Huebl},
  booktitle = {Proceedings of the Platform for Advanced Scientific Computing Conference},
  year = {2024},
  isbn = {979-8-4007-0639-4},
  publisher = {Association for Computing Machinery},
  address = {New York, NY, USA},
  url = {https://doi.org/10.1145/3659914.3659937},
  doi = {10.1145/3659914.3659937},
  articleno = {23},
  numpages = {11},
  location = {Zurich, Switzerland},
  series = {PASC '24}
}

@article{Djordjevi2023TransferLA,
    title = {Transfer learning and multi-fidelity modeling of laser-driven particle acceleration},
    author = {B. Z. Djordjevi{\'c} and J. Kim and Scott C. Wilks and J. D. Ludwig and Conner Myers and Andreas J. Kemp and K. K. Swanson and G. Zeraouli and Elizabeth S. Grace and R. A. Simpson and D. R. Rusby and A. F. Antoine and Peer-Timo Bremer and Jayaraman J. Thiagarajan and Rushil Anirudh and G. J. Williams and T. Ma and MD. A. Mariscal},
    journal = {Physics of Plasmas},
    shortjournal={Phys. Plasmas},
    year = {2023},
    volume = {30},
    number = {4},
    pages = {043111},
    month = {04},
    issn = {1070-664X},
    doi = {10.1063/5.0139285},
    url = {https://doi.org/10.1063/5.0139285},
    eprint = {https://pubs.aip.org/aip/pop/article-pdf/doi/10.1063/5.0139285/17138488/043111\_1\_5.0139285.pdf},
}

@article{adam,
      title={Adam: A Method for Stochastic Optimization}, 
      author={Diederik P. Kingma and Jimmy Ba},
      year={2017},
      eprint={1412.6980},
      archivePrefix={arXiv},
      primaryClass={cs.LG},
      url={https://arxiv.org/abs/1412.6980}, 
}

@inproceedings{optuna,
  title = {Optuna: A Next-generation Hyperparameter Optimization Framework},
  author = {Akiba, Takuya and Sano, Shotaro and Yanase, Toshihiko and Ohta, Takeru and Koyama, Masanori},
  year = {2019},
  isbn = {9781450362016},
  publisher = {Association for Computing Machinery},
  address = {New York, NY, USA},
  url = {https://doi.org/10.1145/3292500.3330701},
  doi = {10.1145/3292500.3330701},
  booktitle = {Proceedings of the 25th {ACM} {SIGKDD} International Conference on Knowledge Discovery and Data Mining},
  pages = {2623–2631},
  numpages = {9},
  location = {Anchorage, AK, USA},
  series = {KDD '19}
}

@inproceedings{asha,
 author = {Li, Liam and Jamieson, Kevin and Rostamizadeh, Afshin and Gonina, Ekaterina and Ben-tzur, Jonathan and Hardt, Moritz and Recht, Benjamin and Talwalkar, Ameet},
 booktitle = {Proceedings of Machine Learning and Systems},
 editor = {I. Dhillon and D. Papailiopoulos and V. Sze},
 pages = {230--246},
 title = {A System for Massively Parallel Hyperparameter Tuning},
 url = {https://proceedings.mlsys.org/paper_files/paper/2020/hash/a06f20b349c6cf09a6b171c71b88bbfc-Abstract.html},
 volume = {2},
 year = {2020}
}

@article{layernormalization,
      title={Layer Normalization}, 
      author={Jimmy Lei Ba and Jamie Ryan Kiros and Geoffrey E. Hinton},
      year={2016},
      eprint={1607.06450},
      archivePrefix={arXiv},
      primaryClass={stat.ML},
      url={https://arxiv.org/abs/1607.06450}, 
}

@article{message_passing,
      title={Neural Message Passing for Quantum Chemistry}, 
      author={Justin Gilmer and Samuel S. Schoenholz and Patrick F. Riley and Oriol Vinyals and George E. Dahl},
      year={2017},
      eprint={1704.01212},
      archivePrefix={arXiv},
      primaryClass={cs.LG},
      url={https://arxiv.org/abs/1704.01212}, 
}

@inproceedings{tpe,
 title = {Algorithms for Hyper-Parameter Optimization},
 author = {Bergstra, James and Bardenet, R\'{e}mi and Bengio, Yoshua and K\'{e}gl, Bal\'{a}zs},
 booktitle = {Advances in Neural Information Processing Systems},
 editor = {J. Shawe-Taylor and R. Zemel and P. Bartlett and F. Pereira and K.Q. Weinberger},
 pages = {},
 publisher = {Curran Associates, Inc.},
 url = {https://papers.nips.cc/paper_files/paper/2011/hash/86e8f7ab32cfd12577bc2619bc635690-Abstract.html},
 volume = {24},
 year = {2011}
}

@article{tune,
      title={Tune: A Research Platform for Distributed Model Selection and Training}, 
      author={Richard Liaw and Eric Liang and Robert Nishihara and Philipp Moritz and Joseph E. Gonzalez and Ion Stoica},
      year={2018},
      eprint={1807.05118},
      archivePrefix={arXiv},
      primaryClass={cs.LG},
      url={https://arxiv.org/abs/1807.05118}, 
}

@inproceedings{oversquashing,
  title={Understanding over-squashing and bottlenecks on graphs via curvature},
  author={Jake Topping and Francesco Di Giovanni and Benjamin Paul Chamberlain and Xiaowen Dong and Michael M. Bronstein},
  booktitle={International Conference on Learning Representations},
  year={2022},
  url={https://openreview.net/forum?id=7UmjRGzp-A}
}

@article{oversmoothing,
      title={A Survey on Oversmoothing in Graph Neural Networks}, 
      author={T. Konstantin Rusch and Michael M. Bronstein and Siddhartha Mishra},
      year={2023},
      eprint={2303.10993},
      archivePrefix={arXiv},
      primaryClass={cs.LG},
      url={https://arxiv.org/abs/2303.10993}, 
}

@article{plasma_physics_discovery, 
    title={Unsupervised discovery of nonlinear plasma physics using differentiable kinetic simulations}, 
    volume={88}, 
    DOI={10.1017/S0022377822000939}, 
    number={6},
    journal={Journal of Plasma Physics},
    shortjournal={J. Plasma Phys.},
    author={Joglekar, Archis S. and Thomas, Alexander G.R.}, 
    year={2022}, 
    pages={905880608}
}

@inproceedings{discovering_symbolic_models_2020,
 author = {Cranmer, Miles and Sanchez Gonzalez, Alvaro and Battaglia, Peter and Xu, Rui and Cranmer, Kyle and Spergel, David and Ho, Shirley},
 booktitle = {Advances in Neural Information Processing Systems},
 editor = {H. Larochelle and M. Ranzato and R. Hadsell and M.F. Balcan and H. Lin},
 pages = {17429--17442},
 publisher = {Curran Associates, Inc.},
 title = {Discovering Symbolic Models from Deep Learning with Inductive Biases},
 url = {https://proceedings.neurips.cc/paper/2020/hash/c9f2f917078bd2db12f23c3b413d9cba-Abstract.html},
 volume = {33},
 year = {2020}
}

@article{rediscovering_orbital_mechanics_2023,
    doi = {10.1088/2632-2153/acfa63},
    url = {https://dx.doi.org/10.1088/2632-2153/acfa63},
    year = {2023},
    month = {10},
    publisher = {IOP Publishing},
    volume = {4},
    number = {4},
    pages = {045002},
    author = {Lemos, Pablo and Jeffrey, Niall and Cranmer, Miles and Ho, Shirley and Battaglia, Peter},
    title = {Rediscovering orbital mechanics with machine learning},
    journal = {Machine Learning: Science and Technology},
    shortjournal = {Mach. Learn.: Sci. Technol.}
}

@article{SavitzkyGolayFilter,
author = {Savitzky, Abraham. and Golay, M. J. E.},
title = {Smoothing and Differentiation of Data by Simplified Least Squares Procedures},
journal = {Analytical Chemistry},
shortjournal = {Anal. Chem.},
volume = {36},
number = {8},
pages = {1627-1639},
year = {1964},
doi = {10.1021/ac60214a047},
URL = {https://doi.org/10.1021/ac60214a047},
eprint = {https://doi.org/10.1021/ac60214a047}
}

@article{SavitzkyGolayFilterErrors,
author = {Steinier, Jean. and Termonia, Yves. and Deltour, Jules.},
title = {Smoothing and differentiation of data by simplified least square procedure},
journal = {Analytical Chemistry},
shortjournal = {Anal. Chem.},
volume = {44},
number = {11},
pages = {1906-1909},
year = {1972},
doi = {10.1021/ac60319a045},
note ={PMID: 22324618},
URL = {https://doi.org/10.1021/ac60319a045},
eprint = {https://doi.org/10.1021/ac60319a045}
}

@article{laser_plasma_instabilities_in_indirect_drive_inertial_fusion,
    author = {Montgomery, David S.},
    title = {Two decades of progress in understanding and control of laser plasma instabilities in indirect drive inertial fusion},
    journal = {Physics of Plasmas},
    shortjournal = {Phys. Plasmas},
    volume = {23},
    number = {5},
    pages = {055601},
    year = {2016},
    month = {04},
    issn = {1070-664X},
    doi = {10.1063/1.4946016},
    url = {https://doi.org/10.1063/1.4946016},
    eprint = {https://pubs.aip.org/aip/pop/article-pdf/doi/10.1063/1.4946016/19945097/055601\_1\_1.4946016.pdf},
}

\end{document}